\documentclass[10pt, twocolumn, twoside]{IEEEtran}

\usepackage{pifont}
\usepackage{cite}
\usepackage[dvips]{epsfig}
\usepackage{graphicx}
\usepackage{calligra}
\usepackage[T1]{fontenc}
\usepackage{bbold}
\usepackage{mathrsfs}

\usepackage{subfigure}
\usepackage{color}
\usepackage{amsmath}
\usepackage{cases}

\usepackage{amssymb}
\usepackage[justification=centering]{caption}
\usepackage{framed}
\usepackage{graphicx}
\usepackage{subfigure}

\usepackage{subfigure}
\usepackage{color}
\usepackage{amsmath}
\usepackage{bm}
\usepackage{amssymb}
\usepackage{amsbsy}
\usepackage{bbm}
\usepackage{dsfont}
\usepackage[normalem]{ulem}
\usepackage{xcolor,cancel}
\usepackage{framed}
\usepackage{cancel}

\usepackage{url}

\colorlet{shadecolor}{yellow}

\usepackage{soul}
\soulregister\cite{7}
\soulregister\ref{7}
\soulregister\pageref{7}

\newtheorem{theorem}{Theorem}

\newtheorem{corollary}{Corollary}

\newtheorem{assumption}{Assumption}
\newtheorem{definition}{Definition}
\ifCLASSINFOpdf
\else
\fi
\begin{document}
%
\title{
Functional Forms of Optimum Spoofing Attacks for Vector Parameter Estimation in Quantized Sensor Networks}
%
%
%

\author{{Jiangfan Zhang,~\IEEEmembership{Student Member,~IEEE},
        Rick S. Blum,~\IEEEmembership{Fellow,~IEEE}, Lance Kaplan,~\IEEEmembership{Fellow,~IEEE}, and Xuanxuan Lu}


\thanks{This work was supported by the U. S. Army Research Laboratory and the U. S. Army Research Office and was accomplished under Agreement Numbers W911NF-14-1-0245 and W911NF-14-1-0261. The views and conclusions contained in this document are those of the authors and should not be interpreted as representing the official policies, either expressed or implied, of the Army Research Laboratory, Army Research Office, or the U.S. Government. The U.S. Government is authorized to reproduce and distribute reprints for Government purposes notwithstanding any copyright notation here on.}


}


\maketitle

\begin{abstract}

Estimation of an unknown deterministic vector from quantized sensor data is considered in the presence of spoofing attacks which alter the data presented to several sensors.  Contrary to previous work, a generalized attack model is employed which manipulates the data using transformations with arbitrary functional forms determined by some attack parameters whose values are unknown to the attacked system.  For the first time, necessary and sufficient conditions are provided under which the transformations provide a guaranteed attack performance in terms of Cramer-Rao Bound (CRB) regardless of the processing the estimation system employs, thus defining a highly desirable attack.  Interestingly, these conditions imply that, for any such attack when the attacked sensors can be perfectly identified by the estimation system, either the Fisher Information Matrix (FIM) for jointly estimating the desired and attack parameters is singular or that the attacked system is unable to improve the CRB for the desired vector parameter through this joint estimation even though the joint FIM is nonsingular.  It is shown that it is always possible to construct such a highly desirable attack by properly employing a sufficiently large dimension attack vector parameter relative to the number of quantization levels employed, which was not observed previously. To illustrate the theory in a concrete way, we also provide some numerical results which corroborate that under the highly desirable attack, attacked data is not useful in reducing the CRB.   

\end{abstract}


%

\section{Introduction}
\label{Section_Introduction}

Recent developments in sensor technology have encouraged a large number of applications of sensor networks for parameter estimation ranging from inexpensive commercial systems to complex military and homeland defense surveillance systems \cite{akyildiz2002survey}. Typically, large-scale sensor networks are comprised of low-cost and spatially distributed sensor nodes with limited battery power and low computing capacity, which makes the system vulnerable to cyberattacks by adversaries. This has led to great interest in studying the vulnerability of sensor networks in various applications and from different perspectives, see \cite{li2005robust, Lee2012Characterization, cui2012coordinated, vempaty2013distributed, nadendla2014distributed, zhang2015Asymptotically, alnajjab2015attacks, zhang2015distributed, niu2015false} and the references therein.  Due to the dominance of digital technology, a great deal of attention has focused on parameter estimation using quantized data \cite{papadopoulos2001sequential, xiao2006distributed2, ribeiro2006bandwidth1, niu2006target, fang2009hyperplane}. 
The sequel considers the problem of estimating a vector parameter by using quantized data collected from a distributed sensor network under the assumption that the measurements from several subsets of sensors have been falsified by spoofing attacks, a topic that has received virtually no attention to date. To be specific, the spoofing attacks maliciously modify the temporal analog measurements of the phenomenon acquired at the subset of attacked sensors. 


\subsection{System and Adversary Models}
\label{SEC:Model}


Consider a distributed sensor network ${\cal S}_N = \{1,2,...,N\}$ consisting of $N$ spatially distributed sensors, with each making some measurements of a particular phenomenon. We assume that the $j$-th sensor acquires $K_j$ measurements, and we denote the before-attack measurement of the $j$-th sensor at time instant $k$ by $x_{jk}$ which follows a probability density function (pdf) ${f_{jk}}\left( {{x_{jk}}\left| {\boldsymbol{\theta }} \right.} \right)$ depending on an unknown deterministic vector parameter ${\boldsymbol{\theta }}$ with dimension $D_{{\boldsymbol{\theta }}}$ that is to be estimated from the measurements. For simplicity, 
we assume that the measurements $\{x_{jk}\}$ are statistically independent but not necessarily identically distributed.

\begin{figure}[htb]
	\vspace{-0.1in}
	\centerline{
		\includegraphics[width=0.45\textwidth]{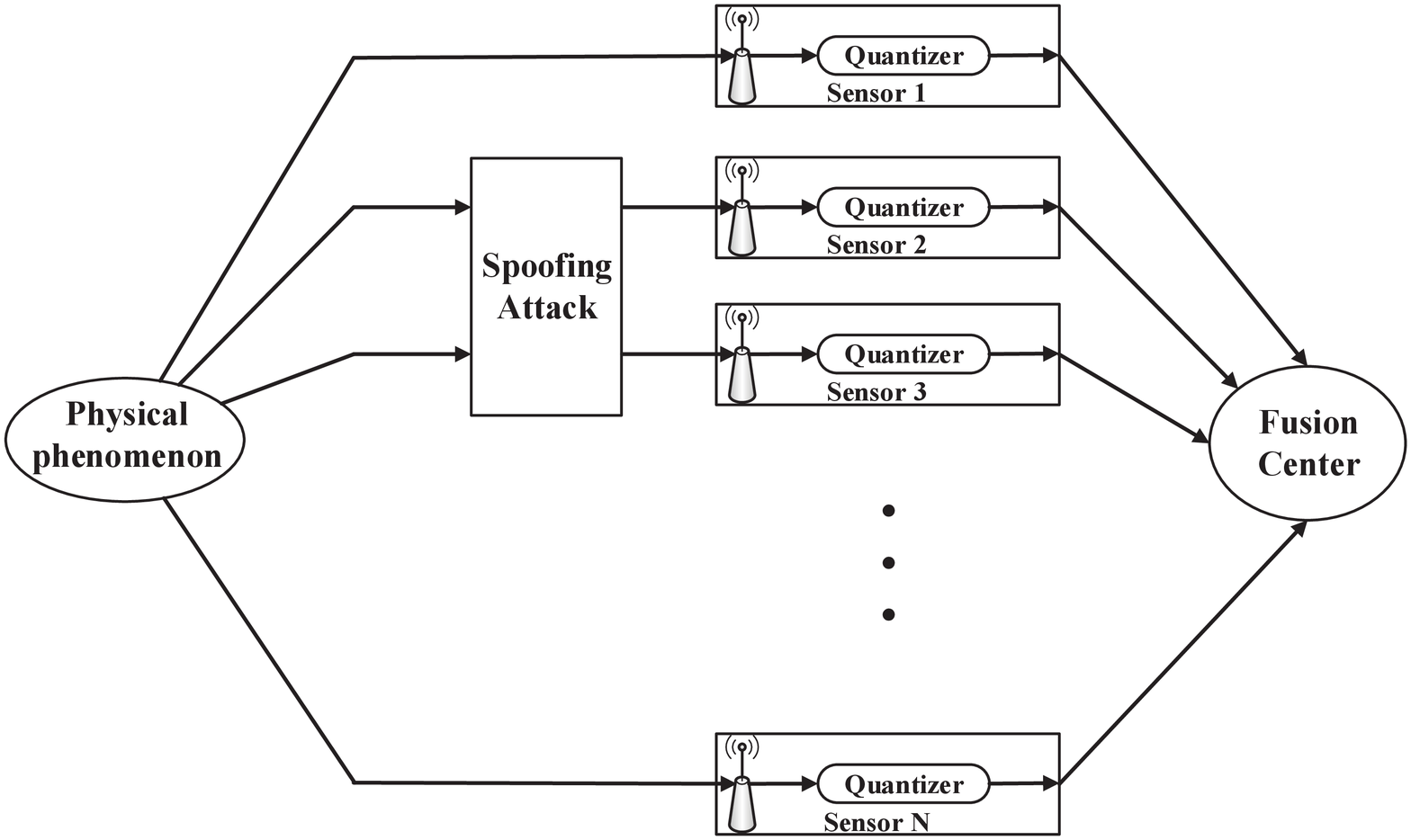}
	}
	\caption{Distributed Estimation System in the Presence of Spoofing Attacks.}
	\label{Fig_System_model}
\end{figure}

The adversaries alter the physical phenomenon as in Fig. \ref{Fig_System_model}, thus
tampering with the measurements at a subset of sensors in the sensor network, hoping to undermine the estimation performance of the system. 
Let ${\cal V} \subset {\cal S}_N$ denote the set of sensors undergoing spoofing attacks while the set ${{{\cal U}}} \buildrel \Delta \over = {{\cal S}_N}\backslash {\cal V}$ represents the set of unattacked sensors.  A generalized mathematical model of spoofing attacks which maliciously modify the distribution of the analog observations of the physical phenomenon at the attacked sensors is considered employing general probability density functions $\{f_{jk}\}$ and $\{g_{jk}\}$ which depend on the desired and attack vector parameters.  To conform to previous work, the functional forms of the attacks, thus $\{f_{jk}\}$ and $\{g_{jk}\}$, are assumed known to the attacked system but the desired and attack vector parameters are not. Thus, the after-attack version ${{\tilde x}_{jk}}$ of $x_{jk}$ obeys the statistical model that\footnote{The notations ${{{\tilde x}_{jk}}}$ and ${{{\tilde u}_{jk}}}$ denote the after-attack analog measurements and the corresponding quantized measurements.} $\{{{\tilde x}_{jk}}\}$ is independent and
\begin{equation}
{{\tilde x}_{jk}} \sim \left\{ \begin{array}{l}
{f_{jk}}\left( {{\tilde x_{jk}}\left| {\boldsymbol{\theta }} \right.} \right), \quad {\text{if }} j \in {{{\cal U}}} \\
{g_{jk}}\left( {{\tilde x_{jk}}\left| {{\boldsymbol{\theta }},{{\boldsymbol{\xi  }}^{(j)}}} \right.} \right), \; {\text{if }} j \in {{\cal V}}
\end{array} \right.,
\end{equation}
where if $j \in {\cal V}$, the after-attack pdf ${g_{{jk}}}( {{x_{jk}} | {{\boldsymbol{\theta }},{{\boldsymbol{\xi  }}^{(j)}}} } )$ is parameterized by the desired vector parameter ${\boldsymbol{\theta }}$ and the attack vector parameter ${{\boldsymbol{\xi  }}^{(j)}}$. It is worth mentioning that the notation
${g_{{jk}}}( {{x_{jk}} | {{\boldsymbol{\theta }},{{\boldsymbol{\xi  }}^{(j)}}} } )$ does not imply that the after-attack pdf ${g_{{jk}}}( {{x_{jk}} | {{\boldsymbol{\theta }},{{\boldsymbol{\xi  }}^{(j)}}} } )$ of the measurements at the $j$-th sensor has to depend on $\boldsymbol{\theta}$.
For example, the adversaries can intercept the signal from the physical phenomenon and generate a new signal using some different pdf solely based on its attack vector parameter.
A detailed example of a practical attack of the type described in (1) is provided in Section~II. 

The set $\cal V$ of attacked sensors can be divided into disjoint subsets $\left\{ {{{\cal A}_p}} \right\}_{p = 1}^P$
in terms of distinct attack vector parameters $\{ {\boldsymbol{\xi}}^{(j)} \}$ such that 
\begin{equation}
{\cal V} = \mathop  \cup \limits_{p = 1}^P {{\cal A}_p}, \; \text{ and } \; {{\cal A}_l} \cap {{\cal A}_m} = \emptyset, \; \forall l \ne m,
\end{equation}
where the attacked sensors in the subset ${\cal A}_p$ are known by the system under attack to employ an identical attack vector parameter ${\boldsymbol{\tau}}^{(p)}$ with dimension $D_p$ so that ${{\boldsymbol{\xi }}^{(j)}} = {{\boldsymbol{\tau }}^{(p)}}$, $\forall j \in {\cal A}_p$.  
The identical attack vectors are possibly due to the sensors in ${\cal A}_p$ being attacked by the same attacker. For the sake of notational simplicity, we use ${\cal A}_0$ to denote the set ${{\cal U}}$ of unattacked sensors.

Due to the communications employed, each sensor is restricted to convert analog measurements to digital data before transmitting this data to the 
fusion center (FC) as shown in Fig. \ref{Fig_System_model}. At the $j$-th sensor, each after-attack measurement ${\tilde{x}}_{jk}$ is quantized to ${\tilde{u}}_{jk}$
by using a $R_j$-level quantizer with quantization regions $\{ I_j^{(r)}\} _{r = 1}^{{R_j}}$, that is,
\begin{equation}
{{\tilde u}_{jk}} = \sum\limits_{r = 1}^{{R_j}} {r {\mathbbm{1}}\left\{ {{{\tilde x}_{jk}} \in I_j^{(r)}} \right\}},
\end{equation}
where ${\mathbbm{1}}\{ \cdot \}$ is the indicator function.
We adopt this general quantization model due to the fact that optimized quantization regions $\{ I_j^{(r)}\} _{r = 1}^{{R_j}}$ for different sensors can be very different, since the measurements from different sensors do not necessarily obey an identical pdf \cite{ribeiro2006bandwidth1, venkitasubramaniam2007quantization}.   
We assume that the quantizer design $\{ I_j^{(r)}\} _{r = 1}^{{R_j}}$ for each sensor is predefined and known to the FC, but not the attacker. 

Let $\boldsymbol{{\Theta}} $ denote a vector containing the unknown parameter ${\boldsymbol{\theta }}$ along with all the unknown attack vector parameters which parameterize the spoofing attacks in the sensor network
\begin{equation} \label{Theta_def}
{\boldsymbol{{\Theta}}}  \buildrel \Delta \over = {\left[ {{{\boldsymbol{\theta }}^T},{{\left( {{{\boldsymbol{\tau }}^{(1)}}} \right)}^T},...,{{\left( {{{\boldsymbol{\tau }}^{(P)}}} \right)}^T}} \right]^T}.
\end{equation}
For the sake of notational simplicity in the following parts, we use $p_{jr}^{(k)}$ to denote the after-attack probability mass function (pmf) of the quantized measurement ${{\tilde u}_{jk}}$ evaluated at ${{\tilde u}_{jk}} =r$, that is,
\begin{align} \notag
p_{jr}^{(k)}\! & \buildrel \Delta \over =\! \Pr \left( {{{\tilde u}_{jk}} = r\left| {\boldsymbol{{\Theta}}}  \right.} \right) \\ \label{pjr}
& \!= \! \left\{ \begin{array}{l}
\!\!\! \int_{I_j^{(r)}}  {{f_{jk}}\left( {\tilde x_{jk} \! \left| {\boldsymbol{\theta }} \right.} \right)d\tilde x_{jk} },  \forall j \in {{{\cal A}_0}}\\
\!\!\! \int_{I_j^{(r)}}  {{g_{{jk}}}\left( {\tilde x_{jk} \! \left| {{\boldsymbol{\theta }},{{\boldsymbol{\tau }}^{(p)}}} \right.} \right)d\tilde x_{jk}},   \forall j \in {\cal A}_p,  \forall p \ge 1
\end{array} \right..
\end{align}
For simplicity, the communication channel between the FC and each sensor is assumed ideal, and hence the FC is able to accurately receive what was transmitted from both the unattacked and attacked sensors. After receiving the quantized data from all sensors, the FC attempts to make an estimate of the desired vector parameter without knowledge of which sensors have been attacked nor the attack parameters used by the attackers.

\subsection{Performance Metric}
\label{Section_Performance_metric}

It is of considerable interest to investigate the performance of spoofing attacks, 
and mathematically characterize the class of the most devastating spoofing attacks under the assumption that the adversaries have no information about what computations the FC is using.
This paper develops guarantees for the attacker's performance that are independent of the computations performed at the FC.
It is clear that if the FC has the information about the groupings of similarly attacked sensors, i.e., $\{{\cal A}_p\}$,
it can use this information to improve estimation performance over the case where this information is not employed. The FC can always do better in estimating the desired vector parameter with extra knowledge. Therefore, for spoofing attacks employing some specific $\{{f_{jk}}( {{x_{jk}}| \boldsymbol{\theta} } )\}$ and $\{{g_{jk}}( {{\tilde x_{jk}}| \boldsymbol{\theta}, {\boldsymbol{\tau}}^{(p)} } )\}$, the case where the compromised sensors are correctly categorized into $P$ different groups according to distinct types of attacks corresponds to the case where the FC has the best chance to combat the spoofing attacks. 
In other words, the best possible estimation performance (smallest error) under this case provides a lower bound on the estimation performance for any other cases, which implies that the corresponding spoofing attack performance under this case provides a guaranteed attack performance in degrading the estimation performance no matter what computations the FC is using.
The recent work in \cite{alnajjab2015attacks} has shown that for some classes of spoofing attacks, with a sufficient number of  observations, the FC is able to perfectly identify the set of unattacked sensors and categorize the attacked sensors into different groups according to distinct types of spoofing attacks. For these reasons, we adopt the following definition of the optimal guaranteed degradation spoofing attacks in this paper.  
\begin{definition} \label{Definition_Opt_Measurement_Attacks}
Consider attacks employing  $\{{f_{jk}}( {{x_{jk}}| \boldsymbol{\theta} } )\}$ and $\{{g_{jk}}( {{\tilde x_{jk}}| \boldsymbol{\theta}, {\boldsymbol{\tau}}^{(p)} } )\}$.  The optimal guaranteed degradation spoofing attack (OGDSA) maximizes the degradation\footnote{See (\ref{Upper_bound_J_theta_estimable}) for example.} of the Cramer-Rao Bound (CRB) for the vector parameter of interest at the FC when the attacked sensors are well identified and categorized according to distinct types of spoofing attacks by the FC. 
\end{definition}

The estimation performance for a vector parameter in a distributed sensor network can be expressed using an error correlation matrix. However, in most cases, a closed form expression for the error correlation matrix is intractable.  Thus the CRB, an asymptotically achievable lower bound on the error correlation matrix, is employed in \emph{Definition \ref{Definition_Opt_Measurement_Attacks}}. 
It is worth mentioning that the optimal guaranteed degradation spoofing attack defined in \emph{Definition \ref{Definition_Opt_Measurement_Attacks}} achieves the classical definition of attack optimality (largest CRB) for the scenario where the FC has the best chance to combat the spoofing attacks. It might not be the classically optimal spoofing attack for the scenario where the FC is unable to determine which sensors are attacked, or to classify sensors into groups of distinct types of spoofing attacks. However, the OGDSAs defined in \emph{Definition \ref{Definition_Opt_Measurement_Attacks}} can provide a guarantee that the actual degradation in the CRB must exceed some critical value no matter what computations the estimation system employs. This guarantee makes OGDSA an excellent spoofing attack from the adversary's point of view.

\subsection{Summary of Results and Main Contributions}

Unlike previous work, a generalized attack model is employed which manipulates the data using transformations with arbitrary functional forms determined by some attack parameters whose values are unknown to the attacked system.  For the first time, necessary and sufficient conditions are provided under which these transformations provide an OGDSA.   
These conditions imply that, for an OGDSA, either the Fisher Information Matrix (FIM) under the conditions of \emph{Definition \ref{Definition_Opt_Measurement_Attacks}} for jointly estimating the desired and attack parameters is singular or that the attacked system is unable to improve the CRB under the conditions of \emph{Definition \ref{Definition_Opt_Measurement_Attacks}} for the desired vector parameter through this joint estimation even though the joint FIM is nonsingular. 
It is shown that when the number of temporal measurements at each sensor is given, it is always possible to construct an OGDSA by properly employing a sufficiently large dimension attack vector parameter relative to the number of quantization levels employed, which was not observed previously.
It is shown that a spoofing attack can render the attacked measurements useless in terms of reducing the CRB under the conditions of \emph{Definition \ref{Definition_Opt_Measurement_Attacks}} for estimating the desired vector parameter if and only if it is an OGDSA.  None of these contributions are provided in the previous work.

In order to illustrate the theory just described in a concrete way, we also provide some numerical results.  For a specific class of OGDSAs, an enhanced  Expectation-Maximization-based algorithm that attempts to use all the attacked and unattacked data to jointly estimate the desired and attack parameters is shown, for a sufficient number of observations, to essentially achieve the CRB which knows which sensors are attacked and only uses data from unattacked sensors.
For completeness, we specify the Expectation-Maximization-based algorithm for general attacks and enhance it with a heuristic rounding approach previously suggested by others in a different application which seems to significantly improve the Expectation-Maximization-based algorithm. 
The purpose of the algorithm and numerical results is to illustrate the properties of OGDSAs. The numerical results demonstrate that a representative algorithm which tries to use the attacked data is not able to obtain performance that is better than the best achievable performance of an approach that ignores the attacked data. 


\subsection{Related Work}

In recent years, estimation problems under different attacks have seen great interest in various engineering applications, see \cite{li2005robust, Lee2012Characterization, cui2012coordinated, vempaty2013distributed, nadendla2014distributed, zhang2015Asymptotically, alnajjab2015attacks, zhang2015distributed, niu2015false, Roome1990Digital, liu2008attack, kim2011strategic, kosut2011malicious} and the references therein. Rather than the man-in-the-middle attacks which falsify the data transmitted from the sensors to the FC \cite{vempaty2013distributed, nadendla2014distributed, zhang2015Asymptotically}, we are primarily interested in spoofing attacks in this paper, which maliciously modify the measurements of the physical phenomenon at a subset of sensors, see Fig. \ref{Fig_System_model}. 

Spoofing attacks have been widely considered in wireless sensor networks, smart grids, radar systems and sonar systems \cite{li2005robust, Lee2012Characterization, alnajjab2015attacks, niu2015false, Roome1990Digital, kim2003robust, liu2008attack, kim2011strategic, kosut2011malicious,  cui2012coordinated}. Each of these recent works takes one specific type of spoofing attack into account, and investigates the attack performance or the estimation performance. In this paper, we don't focus on one specific type of spoofing attack. Instead, we consider a generalized attack model which can describe the different kinds of spoofing attacks employed in all recent work, and moreover, we make use of this generalized model to provide uniform tools to test if a spoofing attack is optimal in our defined sense. In \cite{kim2011strategic} and \cite{kosut2011malicious}, the authors only considered one specific functional form of the spoofing attacks, the so-called data-injection attacks, so they do not address which functional forms are optimum.  Further, the work in \cite{kim2011strategic} and \cite{kosut2011malicious} is only for smart grid systems, while our work is very general.

Another difference between our work and the other recent work on spoofing attacks in \cite{li2005robust, Lee2012Characterization, Roome1990Digital, niu2015false, kim2003robust, liu2008attack, kim2011strategic, kosut2011malicious,  cui2012coordinated} is that we consider estimation based on quantized data which is typically the case in practice.  Interestingly, we show that the quantization limits the capability of the estimation system to combat the spoofing attacks.  In particular, it is shown that the adversaries can launch a class of quantization induced OGDSAs which are easily constructed in practice. 


\subsection{Notation and Organization}

Throughout this paper, bold upper case letters and bold lower case letters are used to denote matrices and column vectors respectively.
The symbol 
${\mathbbm{1}}\{\cdot\}$ stands for the indicator function. Let ${\left[ {\bf{A}} \right]_{i,j}}$ denote the element in the $i$-th row and $j$-th column of the matrix $\bf{A}$, and ${\mathscr{R}}(\bf A)$ represents the range space of $\bf A$. ${\bf{A}} \succ 0$ and ${\bf{A}} \succeq 0$ imply that the matrix is positive definite and positive semidefinite respectively. 
To avoid cumbersome sub-matrix and sub-vector expressions in this paper, we introduce the following notation. 
The notation ${[{\bf{A}}]}_{{\cal S},:}$ stands for the sub-matrix of ${\bf{A}}$ which consists of the elements with row indices in the set ${\cal S}$, and ${[{\bf{A}}]}_{1:N}$ represents the $N$-by-$N$ leading principle minor of ${\bf{A}}$. 
The $i$-th element of the vector ${\bf{v}}$ is denoted by $v_i$, and ${[ {\bf v} ]}_{{\cal S}}$ represents the sub-vector of $\bf v$ which only contains the elements with indices in the set ${\cal S}$. The symbols ${\nabla _{\bf{v}}}f$ and ${\nabla _{\bf{v}}^2} f$ respectively signify the gradient and Hessian of $f$ with respect to $\bf v$.
Finally, the expectation and rank operators are denoted by ${\mathbbm{E}}\left(  \cdot  \right)$ and ${{\rm{rank}(\cdot)}}$ respectively.

The remainder of the paper is organized as follows. Some illustrative example of a practical spoofing attack is introduced in \emph{Section \ref{SEC:Example}}. \emph{Section \ref{Section_Optimality_Spoofing_Attacks}} provides the necessary and sufficient conditions for the OGDSAs. A joint attack identification and parameter estimation approach is developed in \emph{Section \ref{Sec_JointIdentification}}, which is used in \emph{Section \ref{Section_Numerical_Results}} to corroborate our theoretical results.
Finally, \emph{Section \ref{SEC:Conclusion}} provides our conclusions.


\section{Illustrative Example of a Practical Spoofing Attack}
\label{SEC:Example} 

Spoofing attacks on sensor networks can occur in various engineering applications. For instance, spoofing attacks have been described for the localization problem in wireless sensor networks, see \cite{li2005robust, Lee2012Characterization} and the references therein. \emph{Table I} in \cite{li2005robust} provides a summary of different types of spoofing attack threats for the localization problem. The dangers of spoofing attacks in the Global Positioning System (GPS) that controls everything from car navigation to national power grids have drawn serious public concern \cite{Bland2008GPS, Couts2013Want}. Radar and sonar systems also suffer from spoofing attack threats in practice. As one example of a spoofing attack technique, the application of an electronic countermeasure (ECM), which is designed to jam or deceive the radar or sonar system, can critically degrade the detection and estimation performance of the system \cite{skolnik1980Introduction}. One popular technique for the implementation of ECM employs digital radio frequency memory (DRFM) in radar systems to manipulate the received signal and retransmit it back to confuse the victim radar system. DRFM can mislead the estimation of the range of the target by altering the delay in transmission of pulses, and fool the system into incorrectly estimating the velocity of the target by introducing a Doppler shift in the retransmitted signal \cite{Roome1990Digital}. An example of a spoofing attack created by nature is environmental variation in shallow water sonar systems. According to waveguide-invariant theory \cite{grachev1993theory}, the environmental variation, such as sound-speed or water-depth perturbations, essentially introduces an apparent shift in the position of the target of interest when the data is processed by matched field processing \cite{d1999mirages, kim2003robust}. Hence these environmental variations can be treated as spoofing attacks which falsify the physical model of the received signal in sonar systems. More recently, the data-injection attack considered in smart grids is another typical example of a spoofing attack, see \cite{liu2008attack, kim2011strategic, kosut2011malicious,  cui2012coordinated} and the references therein.

%



In order to motivate the mathematical description of spoofing attacks, we consider a spoofing attack utilizing a DRFM in a radar system as an example, which stores the received signal and strategically retransmits it back by introducing an additional delay to mislead the estimation of the range of the target. In the absence of spoofing attacks, the simplified signal model of the $k$-th measurement $x_{jk}$ at the $j$-th receiver at time instant $t_{jk}$, which ignores the Doppler shift, can be expressed as
\begin{equation} \label{delay_receive_model}
{x_{jk}} = \sqrt {{E_j}} {a_j}s\left( {{t_{jk}} - \theta_j } \right) + {n_{jk}},
\end{equation} 
where $s(\cdot)$, ${E_j}$ and ${a_j}$ respectively represent the transmitted signal, the transmitted energy, and the reflection coefficient. The time delay is denoted by $\theta_j$ which is the parameter to be estimated. We assume that the clutter-plus-noise $n_{jk}$ is a zero-mean Gaussian random variable with known variance ${\sigma}_j^2$ while all other quantities in (\ref{delay_receive_model}) are deterministic. As a result, the pdf ${f_{jk}}( {{x_{jk}}| \theta_j  } )$ of $x_{jk}$ in the absence of spoofing attacks is given by\footnote{${\cal N}\left( a,b \right)$ denotes a Gaussian pdf with mean $a$ and variance $b$.}
\begin{equation}
{f_{jk}}\left( {{x_{jk}}\left| \theta_j  \right.} \right) = {\cal N}\left( {\sqrt {{E_j}} {a_j}s\left( {{t_{jk}} - \theta_j } \right),\sigma _j^2} \right).
\end{equation}
In the presence of a spoofing attack, the after-attack measurement ${\tilde x}_{jk}$ can be described as
\begin{equation} \label{DRFM_after_attack_measurement}
{{\tilde x}_{jk}} = \sqrt {{E_j}} {a_j}s\left( {{t_{jk}} - \theta_j  - \xi_j } \right) + {n_{jk}},
\end{equation}
where $\xi_j$ is the delay introduced by the DRFM. Therefore, we can obtain the corresponding after-attack pdf of ${\tilde x}_{jk}$
\begin{align} \notag
{g_{jk}}  \left( {{\tilde x}_{jk}  \left| {\theta_j  , \xi_j } \right.} \right) & = {\cal N}\left( {\sqrt {{E_j}} {a_j}s\left( {{t_{jk}} - \theta_j - \xi_j } \right),  \sigma _j^2} \right) \\ \label{DRFM_pdf}
& = {f_{jk}}  \left( {{{\tilde x}_{jk}} \left| \theta_j + \xi_j  \right.} \right).
\end{align}
In this example, the after-attack pdf ${g_{jk}}( {{{{{\tilde x}_{jk}}}}| {\theta_j , \xi_j} } )$ and the before-attack pdf ${f_{jk}}( {{{x_{jk}}}| \theta_j  } )$ are in the same family as shown in (\ref{DRFM_pdf}), i.e., the family of Gaussian distributions with the same variance ${\sigma}_j^2$. While this may not always be true, the after-attack pdf is generally not only parameterized by the desired parameter $\theta_j$ but also by an unknown attack parameter $\xi_j$. 

Motivated by this example and other popular spoofing attack examples, such as those in \cite{li2005robust, Lee2012Characterization, alnajjab2015attacks, Roome1990Digital, kim2003robust, liu2008attack, kim2011strategic, kosut2011malicious,  cui2012coordinated}, the essential impact of a spoofing attack at the $j$-th sensor is to maliciously modify the measurements at the $j$-th sensor in a manner similar to (\ref{DRFM_after_attack_measurement}). Hence, any given spoofing attack at the $j$-th sensor can be described as a mapping which maps the before-attack pdfs $\{{f_{jk}}( {{x_{jk}}| \boldsymbol{\theta} } )\}$ of the measurements at the $j$-th sensor to the after-attack pdfs $\{{g_{jk}}( {{x_{jk}}| \boldsymbol{\theta}, {\boldsymbol{\xi}}^{(j)} } )\}$, where $\boldsymbol{\theta}$ and ${\boldsymbol{\xi}}^{(j)}$ account for the desired vector parameter and the attack vector parameter at the $j$-th sensor which represents those deterministic unknowns which can determine the after-attack pdfs. 

\section{The Optimality of spoofing attacks}
\label{Section_Optimality_Spoofing_Attacks}

In this section, we pursue the explicit characterization of the optimal spoofing attack as per \emph{Definition \ref{Definition_Opt_Measurement_Attacks}}. 
The adversaries can attempt to maximize the ${\text{CRB}}$ for $\boldsymbol{\theta}$
to achieve an optimal spoofing attack as per \emph{Definition \ref{Definition_Opt_Measurement_Attacks}}. Hence, we first formulate the FIM for estimating ${\boldsymbol{\Theta}}$ in the following, and then based on the expression of the FIM, we provide the necessary and sufficient conditions for the optimal spoofing attack as per \emph{Definition \ref{Definition_Opt_Measurement_Attacks}}.

The FIM ${{\bf{J}}_{\boldsymbol{{\boldsymbol{\Theta}}}} }$ for estimating ${\boldsymbol{\Theta}}$ is defined as \cite{kay1993fundamentals} 
\begin{equation} \label{FIM_def}
{\left[ {{{\bf{J}}_{\boldsymbol{\Theta}} }} \right]_{l,m}} \buildrel \Delta \over =  - {\mathbbm{E}}\left\{ {\frac{{{\partial ^2}L\left( {{\bf{\Theta }}} \right)}}{{\partial {{{\Theta}} _l}\partial {{{\Theta}} _m}}}} \right\},
\end{equation}
where $L\left( {{\bf{\Theta }}} \right)$ denotes the log-likelihood function.

When the attacked senors are well identified and categorized into different groups according to distinct types of spoofing attacks, the log-likelihood function $L\left( {{\bf{\Theta }}} \right)$ in (\ref{FIM_def}) evaluated at 
\begin{equation} \notag
{\bf{\tilde{u}}} \! \buildrel \Delta \over =\! [{{\tilde{u}}_{11}}, {{\tilde{u}}_{12}},...,{{\tilde{u}}_{1K_1}}, {{\tilde{u}}_{21}},...,{{\tilde{u}}_{NK_N}} ]^T \!= \! {\bf{r}}
\end{equation}
can be expressed as\footnote{Note that if ${p_{jr}^{(k)}} =0$ for some $j$, $k$ and $r$, then we just need to eliminate the corresponding summand in (\ref{LLR}). Hence, without loss of generality, we assume ${p_{jr}^{(k)}} >0$.}
\begin{align}  \notag
L\left( {{\bf{\Theta }}} \right) & = \ln \Pr \left( {{\bf{\tilde u}} = {\bf{r}}\left| {\boldsymbol{\Theta}}  \right.} \right)\\  \label{LLR}
&  = \sum\limits_{p = 0}^P {\sum\limits_{j \in {{\cal A}_p}} {\sum\limits_{k = 1}^{{K_j}} {\sum\limits_{r = 1}^{{R_j}} {{\mathbbm{1}}\left\{ {{r_{jk}} = r} \right\}\ln p_{jr}^{(k)}} } } } 
\end{align}
by employing (\ref{pjr}).

By substituting the expression of the log-likelihood function $L\left( {{\bf{\Theta }}} \right)$ in (\ref{LLR}) into the definition of the FIM in (\ref{FIM_def}), it can be shown that the FIM ${{{\bf{J}}_{\boldsymbol{\Theta}} }}$ for ${\boldsymbol{\Theta}}$ takes the form 
\begin{equation}\label{FIM_jointly_form}
{{\bf{J}}_{\bf{\Theta }}} \buildrel \Delta \over = \left[ {\begin{array}{*{20}{c}}
{{{\bf{J}}_{\boldsymbol{\theta }}}}&{{{\bf{B}}_1}}&{{{\bf{B}}_2}}& \cdots &{{{\bf{B}}_P}}\\
{{\bf{B}}_1^T}&{{{\bf{J}}_{{{\boldsymbol{\tau }}^{(1)}}}}}&{\bf{0}}& \cdots &{\bf{0}}\\
{{\bf{B}}_2^T}&{\bf{0}}&{{{\bf{J}}_{{{\boldsymbol{\tau }}^{(2)}}}}}& \ddots & \vdots \\
 \vdots & \vdots & \ddots & \ddots &{\bf{0}}\\
{{\bf{B}}_P^T}&{\bf{0}}& \cdots &{\bf{0}}&{{{\bf{J}}_{{{\boldsymbol{\tau }}^{(P)}}}}}
\end{array}} \right]
\end{equation}
where ${{{\bf{J}}_{\boldsymbol{\theta }}}} \in {\mathbb{R}}^{D_{\boldsymbol{\theta}} \times D_{\boldsymbol{\theta}}}$, ${{{\bf{J}}_{{\boldsymbol{\tau }}^{(p)}}}} \in {\mathbb{R}}^{D_{p} \times D_{p}}$, and ${{{\bf{B}}_{p}}} \in {\mathbb{R}}^{D_{\boldsymbol{\theta}} \times D_{p}}$ for all $p=1,2,...,P$. 
Moreover, following from (\ref{Theta_def}), (\ref{FIM_def}) and (\ref{LLR}), we can obtain that $\forall p$
\begin{equation} \label{J_tau_p}
{{\bf{J}}_{{{\boldsymbol{\tau }}^{(p)}}}} = \sum\limits_{j \in {{\cal A}_p}} {\sum\limits_{k = 1}^{{K_j}} {\sum\limits_{r = 1}^{{R_j}} {\frac{1}{{p_{jr}^{(k)}}}\frac{{\partial p_{jr}^{(k)}}}{{\partial {{\boldsymbol{\tau }}^{(p)}}}}{{\left[ {\frac{{\partial p_{jr}^{(k)}}}{{\partial {{\boldsymbol{\tau }}^{(p)}}}}} \right]}^T}} } } 
\end{equation}
\begin{equation} \label{B_p}
{{\bf{B}}_p} = \sum\limits_{j \in {{\cal A}_p}} {\sum\limits_{k = 1}^{{K_j}} {\sum\limits_{r = 1}^{{R_j}} {\frac{1}{{p_{jr}^{(k)}}}\frac{{\partial p_{jr}^{(k)}}}{{\partial {\boldsymbol{\theta }}}}{{\left[ {\frac{{\partial p_{jr}^{(k)}}}{{\partial {{\boldsymbol{\tau }}^{(p)}}}}} \right]}^T}} } },
\end{equation}
and
\begin{equation} \label{J_theta}
{{\bf{J}}_{\boldsymbol{\theta }}} = \sum\limits_{p = 0}^P {{{\bf{J}}_{{{\cal A}_p}}}}, 
\end{equation}
where ${{\bf{J}}_{{{\cal A}_p}}}$, which is contributed from the measurements observed at the sensors in ${\cal A}_p$, is defined as
\begin{equation} \label{J_A_p}
{{\bf{J}}_{{{\cal A}_p}}} = \sum\limits_{j \in {{\cal A}_p}} {\sum\limits_{k = 1}^{{K_j}} {\sum\limits_{r = 1}^{{R_j}} {\frac{1}{{p_{jr}^{(k)}}}\frac{{\partial p_{jr}^{(k)}}}{{\partial {\boldsymbol{\theta }}}}{{\left[ {\frac{{\partial p_{jr}^{(k)}}}{{\partial {\boldsymbol{\theta }}}}} \right]}^T}} } }. 
\end{equation}

Let ${N_p}$ denote the number of sensors in ${\cal A}_p$, and let $\{j_i^p\}_{i=1}^{{N_p}}$ stand for the indices of the sensors in ${\cal A}_p$. For each $p$, 
we define two matrices 
\begin{align} \notag
{{\bf{\Phi }}_{{{\boldsymbol{\theta }}^{(p)}}}} \! \buildrel \Delta \over =  &  \bigg[ \phi _{j_1^p11}^{{{\boldsymbol{\theta }}^{(p)}}},\phi _{j_1^p12}^{{{\boldsymbol{\theta }}^{(p)}}},...,\phi _{j_1^p1{R_{j_1^p}}}^{{{\boldsymbol{\theta }}^{(p)}}},\phi _{j_1^p21}^{{{\boldsymbol{\theta }}^{(p)}}},...,\phi _{j_1^p2{R_{j_1^p}}}^{{{\boldsymbol{\theta }}^{(p)}}}, \bigg. \\  \label{Phi_theta_p}
& \; \bigg. \phi _{j_1^p31}^{{{\boldsymbol{\theta }}^{(p)}}}....,\phi _{j_1^p{K_{j_1^p}}{R_{j_1^p}}}^{{{\boldsymbol{\theta }}^{(p)}}},\phi _{j_2^p11}^{{{\boldsymbol{\theta }}^{(p)}}},...,\phi _{j_{{N_p}}^p{K_{j_N^p}}{R_{j_{{N_p}}^p}}}^{{{\boldsymbol{\theta }}^{(p)}}} \bigg],
\end{align}
and
\begin{align} \notag
{\bf{\Phi }}_{{\boldsymbol{\tau }}^{(p)}} \! \buildrel \Delta \over =  &  \bigg[ \phi _{j_1^p11}^{{{\boldsymbol{\tau }}^{(p)}}},\phi _{j_1^p12}^{{{\boldsymbol{\tau }}^{(p)}}},...,\phi _{j_1^p1{R_{j_1^p}}}^{{{\boldsymbol{\tau }}^{(p)}}},\phi _{j_1^p21}^{{{\boldsymbol{\tau }}^{(p)}}},...,\phi _{j_1^p2{R_{j_1^p}}}^{{{\boldsymbol{\theta }}^{(p)}}}, \bigg. \\  \label{Phi_tau_p}
& \; \bigg. \phi _{j_1^p31}^{{{\boldsymbol{\tau }}^{(p)}}}....,\phi _{j_1^p{K_{j_1^p}}{R_{j_1^p}}}^{{{\boldsymbol{\tau }}^{(p)}}},\phi _{j_2^p11}^{{{\boldsymbol{\tau }}^{(p)}}},...,\phi _{j_{{N_p}}^p{K_{j_N^p}}{R_{j_{{N_p}}^p}}}^{{{\boldsymbol{\tau }}^{(p)}}} \bigg],
\end{align}
where the vectors ${\boldsymbol{\phi}} _{jkr}^{{{{\boldsymbol{\theta }}^{(p)}}}}$ and ${\boldsymbol{\phi}}_{jkr}^{{{\boldsymbol{\tau }}^{(p)}}}$ in (\ref{Phi_theta_p}) and (\ref{Phi_tau_p}) are given by
\begin{equation} \label{phi_theta_tau_p}
{\boldsymbol{\phi}} _{jkr}^{{{{\boldsymbol{\theta }}^{(p)}}}} \buildrel \Delta \over = {\sqrt {\frac{{{1}}}{{p_{jr}^{(k)}}}}} \frac{{\partial p_{jr}^{(k)}}}{{\partial {\boldsymbol{\theta }}}}  \quad {\text{and}} \quad {\boldsymbol{\phi}}_{jkr}^{{{\boldsymbol{\tau }}^{(p)}}} \buildrel \Delta \over = {\sqrt {\frac{{{1}}}{{p_{jr}^{(k)}}}}} \frac{{\partial p_{jr}^{(k)}}}{{\partial {{\boldsymbol{\tau }}^{(p)}}}}.
\end{equation}

By employing the singular value decomposition of ${{\bf{\Phi }}_{{{\boldsymbol{\theta }}^{(p)}}}}$ and ${\bf{\Phi }}_{{\boldsymbol{\tau }}^{(p)}}$ for all $p$ 
\begin{align}  \label{Phi_SVD}
{{\bf{\Phi }}_{{{\boldsymbol{\tau }}^{(p)}}}} \!\! = \!\! {{\bf{U}}_{{{\boldsymbol{\tau }}^{(p)}}}}{{\bf{\Lambda }}_{{{\boldsymbol{\tau }}^{(p)}}}} \! {\bf{V}}_{{{\boldsymbol{\tau }}^{(p)}}}^T \text{ and } {{\bf{\Phi }}_{{{\boldsymbol{\theta }}^{(p)}}}} \! \! = \! \! {{\bf{U}}_{{{\boldsymbol{\theta }}^{(p)}}}}{{\bf{\Lambda }}_{{{\boldsymbol{\theta }}^{(p)}}}}\! {\bf{V}}_{{{\boldsymbol{\theta }}^{(p)}}}^T,
\end{align}
the expressions of ${{\bf{J}}_{{{\boldsymbol{\tau }}^{(p)}}}}$, ${{\bf{B}}_p}$, and ${{\bf{J}}_{\boldsymbol{\theta }}}$ in (\ref{J_tau_p})--(\ref{J_theta}) can be written in compact forms following
\begin{equation} \label{J_tau_p_matrix}
{{\bf{J}}_{{{\boldsymbol{\tau }}^{(p)}}}} = {{\bf{\Phi }}_{{{\boldsymbol{\tau }}^{(p)}}}}{\bf{\Phi }}_{{{\boldsymbol{\tau }}^{(p)}}}^T = {{\bf{U}}_{{{\boldsymbol{\tau }}^{(p)}}}}{{\bf{\Lambda }}_{{{\boldsymbol{\tau }}^{(p)}}}}{\bf{\Lambda }}_{{{\boldsymbol{\tau }}^{(p)}}}^T{\bf{U}}_{{{\boldsymbol{\tau }}^{(p)}}}^T,
\end{equation}
\begin{equation} \label{B_p_matrix}
{{\bf{B}}_p} = {{\bf{\Phi }}_{{{\boldsymbol{\theta }}^{(p)}}}}{\bf{\Phi }}_{{{\boldsymbol{\tau }}^{(p)}}}^T,
\end{equation}
and
\begin{align} \notag
{{\bf{J}}_{\boldsymbol{\theta }}} & = \sum\limits_{p = 0}^P {{{\bf{J}}_{{{\cal A}_p}}}}  = \sum\limits_{p = 0}^P {{{\bf{\Phi }}_{{{\boldsymbol{\theta }}^{(p)}}}}{\bf{\Phi }}_{{{\boldsymbol{\theta }}^{(p)}}}^T}  \\  \label{J_theta_matrix} 
& = \sum\limits_{p = 0}^P {{{\bf{U}}_{{{\boldsymbol{\theta }}^{(p)}}}}{{\bf{\Lambda }}_{{{\boldsymbol{\theta }}^{(p)}}}}{\bf{\Lambda }}_{{{\boldsymbol{\theta }}^{(p)}}}^T{\bf{U}}_{{{\boldsymbol{\theta }}^{(p)}}}^T}.
\end{align}

\subsection{Inestimable Spoofing Attacks}
\label{Section_Inestimable_Spoofing_Attacks}


Next we show that just due to the sensor system employing a quantization with a limited alphabet, the adversaries can launch a class of spoofing attacks which bring about a singular FIM ${\bf{J}}_{\boldsymbol{\Theta}}$ due to the singularity of ${\bf{J}}_{{\boldsymbol{\tau}}^{(p)}}$ for some $p \in \{1,2,...,P\}$. We formally define these inestimable spoofing attacks as follows.
\begin{definition}[Inestimable spoofing attack] \label{Definition_inestimable_attack}
The $p$-th spoofing attack is referred to as an inestimable spoofing attack (ISA) if the corresponding ${\bf{J}}_{{\boldsymbol{\tau}}^{(p)}}$ defined in (\ref{J_tau_p}) is singular.
\end{definition}

From (\ref{J_tau_p}), we have the following result with regard to the singularity of ${\bf{J}}_{{\boldsymbol{\tau}}^{(p)}}$.
\begin{theorem} \label{Theorem_inestimable_attack}
For the $p$-th spoofing attack, if the dimension $D_p$ of the attack parameter ${\boldsymbol{\tau}}^{(p)}$ satisfies
\begin{equation} \label{Theorem_inestimable_attack_sufficient_condition}
D_p >\sum\limits_{j \in {{\cal A}_p}} {{K_j}\left( {{R_j} - 1} \right)},
\end{equation}
then ${\bf{J}}_{{\boldsymbol{\tau}}^{(p)}}$ is singular, and furthermore, the FIM ${\bf{J}}_{\boldsymbol{\Theta}}$ is also singular. 
\end{theorem}
\begin{IEEEproof}
It is clear that 
\begin{equation}
\sum\limits_{r = 1}^{{R_j}} {p_{jr}^{(k)}}  = 1,
\end{equation}
for all $j$ and $k$. Hence, we can obtain that
\begin{equation}
\sum\limits_{r = 1}^{{R_j}} {\frac{{\partial p_{jr}^{(k)}}}{{\partial { {\boldsymbol{\tau}}^{(p)} }}}}  = {\bf 0}, \; {\forall j  \text{ and } k},
\end{equation}
which yields
\begin{equation}
{\rm{rank}}\left( {\sum\limits_{r = 1}^{{R_j}} {\frac{{\partial p_{jr}^{(k)}}}{{\partial   {\boldsymbol{\tau}}^{(p)}  }}{{\left[ {\frac{{\partial p_{jr}^{(k)}}}{{\partial {\boldsymbol{\tau}}^{(p)}  }}} \right]}^T}} } \right) \le {R_j} - 1, \; {\forall j \text{ and } k}.
\end{equation}

 
Thus, the rank of ${\bf{J}}_{{\boldsymbol{\tau}}^{(p)}}$ is bounded above as per
\begin{align} \notag
{\rm{rank}}\left( {{{\bf{J}}_{{{\boldsymbol{\tau }}^{(p)}}}}} \right) & = {\rm{rank}}\left( {\sum\limits_{j \in {{\cal A}_p}} {\sum\limits_{k = 1}^{{K_j}} {\sum\limits_{r = 1}^{{R_j}} {\frac{1}{{p_{jr}^{(k)}}}\frac{{\partial p_{jr}^{(k)}}}{{\partial {{\boldsymbol{\tau }}^{(p)}}}}{{\left[ {\frac{{\partial p_{jr}^{(k)}}}{{\partial {{\boldsymbol{\tau }}^{(p)}}}}} \right]}^T}} } } } \right)\\ \notag
& \le \sum\limits_{j \in {{\cal A}_p}} {\sum\limits_{k = 1}^{{K_j}} {{\rm{rank}}\left( {\sum\limits_{r = 1}^{{R_j}} {\frac{{\partial p_{jr}^{(k)}}}{{\partial {{\boldsymbol{\tau }}^{(p)}}}}{{\left[ {\frac{{\partial p_{jr}^{(k)}}}{{\partial {{\boldsymbol{\tau }}^{(p)}}}}} \right]}^T}} } \right)} } \\ \label{rank_J_tau_upper_bound}
& \le \sum\limits_{j \in {{\cal A}_p}} {{K_j}\left( {{R_j} - 1} \right)}. 
\end{align}
Since ${\bf{J}}_{{\boldsymbol{\tau}}^{(p)}}$ is a $D_p$-by-$D_p$ positive semidefinite matrix, we know that ${\bf{J}}_{{\boldsymbol{\tau}}^{(p)}}$ is singular if $D_p > \sum\nolimits_{j \in {{\cal A}_p}} {{K_j}\left( {{R_j} - 1} \right)}$. Finally, the proof concludes by noting that ${\bf{J}}_{\boldsymbol{\Theta}}$ is singular as long as ${\bf{J}}_{{\boldsymbol{\tau}}^{(p)}}$ is singular.
\end{IEEEproof}


The proof of \emph{Theorem \ref{Theorem_inestimable_attack}} demonstrates that the rank of ${\bf{J}}_{{\boldsymbol{\tau}}^{(p)}}$ is upper bounded by the sum in (\ref{rank_J_tau_upper_bound})
which is determined by the number of temporal measurements and the size of the alphabet set employed at each sensor under the $p$-th spoofing attack.
This implies that when the number of temporal measurements at each attacked sensor is given, the numbers of quantization levels employed at the compromised sensors will limit the size of the attack vector parameter the quantized estimation system can estimate with an accuracy that increases with more observations.
\emph{Theorem \ref{Theorem_inestimable_attack}} provides a sufficient condition under which inestimable spoofing attacks can be launched.
Thus, these inestimable spoofing attacks, which are quantization induced, can be easily constructed in practice,
even without any information about the value of $\boldsymbol{\theta}$ and the quantization regions $\{I_j^{(r)}\}$ at each sensor. Further, if the adversaries have knowledge of the number of quantization levels of each attacked sensor and the number of temporal measurements at each attacked sensor, they know the minimum size of the attack vector parameter they can employ to ensure an inestimable spoofing attack.
One simple example of an inestimable spoofing attack employs $D_p > \sum\nolimits_{j \in {{\cal A}_p}} {{K_j}\left( {{R_j} - 1} \right)}$ and
\begin{equation}
{{\tilde x}_{{j}k}} = \sum\limits_{i = 1}^{{D_p}} {{\tau}} _i^{(p)}\left({x_{{j}k}}\right)^i.
\end{equation}




If (\ref{Theorem_inestimable_attack_sufficient_condition}) is not satisfied, the inestimability is determined by the $\{I_j^{(r)}\}$ employed at the attacked sensors
and the set of after-attack pdfs $\{ {g_{jk}}( {x_{jk}| {{\boldsymbol{\theta }},{{\boldsymbol{\tau }}^{(p)}}} } ) \}$. 
From (\ref{J_tau_p_matrix}), it is seen that the inestimability of the $p$-th spoofing attack is equivalent to 
\begin{equation} \label{ISA_condition}
{\text{rank}}\left( {{\bf{\Lambda }}_{{{\boldsymbol{\tau }}^{(p)}}}} \right) < {D_p}.
\end{equation} 

In the presence of inestimable spoofing attacks, the FIM ${\bf{J}}_{\boldsymbol{\Theta}}$ for joint estimation of the desired vector parameter and the attack vector parameters is singular, which implies that the FC is unable to improve the estimation of $\boldsymbol{\theta}$ via jointly estimating $\boldsymbol{\theta}$ and the attack vector parameters in the CRB sense.
If (\ref{ISA_condition}) is true for all $p=1,2,...,P$, this means the best the FC can do in this sense is to estimate ${\boldsymbol{\theta}}$ using only unattacked data, and hence the CRB for ${\boldsymbol{\theta}}$ in such case can be obtained as
\begin{equation} \label{CRB_inestimable}
{\text{CRB}}_{\text{ISA}}\left({\boldsymbol{\theta}}\right) = {\bf{J}}_{{{\cal A}_0}}^{ - 1} = {{\bf{U}}_{{{\boldsymbol{\theta }}^{(0)}}}}{\left( {{{\bf{\Lambda }}_{{{\boldsymbol{\theta }}^{(0)}}}}{\bf{\Lambda }}_{{{\boldsymbol{\theta }}^{(0)}}}^T} \right)^{ - 1}}{\bf{U}}_{{{\boldsymbol{\theta }}^{(0)}}}^T
\end{equation}
by employing (\ref{J_theta_matrix}).

\subsection{Optimal Estimable Spoofing Attacks}
\label{Section_Optimal_Estimable_Spoofing_Attacks}

 
In this subsection, we focus on estimable spoofing attacks (defined next), and obtain the necessary and sufficient conditions for the optimal estimable spoofing attacks via FIM analysis.
\begin{definition}[Estimable spoofing attack]
The $p$-th spoofing attack is said to be estimable if the corresponding ${\bf{J}}_{{\boldsymbol{\tau}}^{(p)}}$ defined in (\ref{J_tau_p}) is nonsingular.
\end{definition}

Without loss of generality, we assume all spoofing attacks are estimable in this subsection. Otherwise, we can eliminate 
the observations at ISA sensors,
and just consider the joint estimation of the desired vector parameter ${\boldsymbol{\theta}}$ and the estimable attack vector parameters.

From (\ref{FIM_jointly_form}) and (\ref{J_theta}), we can obtain the CRB for ${\boldsymbol{\theta}}$ in the presence of estimable spoofing attacks as 
\begin{align} \notag
{\left[ {{\bf{J}}_{\boldsymbol{\Theta }}^{ - 1}} \right]_{1:{D_{\boldsymbol{\theta }}}}} &  = {\left( {{{\bf{J}}_{\boldsymbol{\theta }}} - \sum\limits_{p = 1}^P {{{\bf{B}}_p}{\bf{J}}_{{{\boldsymbol{\tau }}^{(p)}}}^{ - 1}{\bf{B}}_p^T} } \right)^{ - 1}} \\ \label{Inverse_FIM_from_identities_Simplified}
& = {\left[ {{{\bf{J}}_{{{\cal A}_0}}} + \sum\limits_{p = 1}^P {\left( {{{\bf{J}}_{{{\cal A}_p}}} - {{\bf{B}}_p}{\bf{J}}_{{{\boldsymbol{\tau }}^{(p)}}}^{ - 1}{\bf{B}}_p^T} \right)} } \right]^{ - 1}}.
\end{align}




In the following theorem, we provide an upper bound on the CRB for ${\boldsymbol{\theta}}$ in (\ref{Inverse_FIM_from_identities_Simplified}) in the positive semidefinite sense.

%
\begin{theorem} \label{Theorem_estimable_FIM_upper_bound}
In the presence of estimable spoofing attacks, the CRB for $\boldsymbol{\theta}$ is bounded above as per
\begin{equation} \label{Upper_bound_J_theta_estimable}
{\text{CRB}}_{\text{ESA}}\left({\boldsymbol{\theta}}\right) \buildrel \Delta \over = {\left[ {{\bf{J}}_{\boldsymbol{\Theta }}^{ - 1}} \right]_{1:{D_{\boldsymbol{\theta }}}}} \preceq {\bf{J}}_{{{\cal A}_0}}^{ - 1}.
\end{equation}
Equality in (\ref{Upper_bound_J_theta_estimable}) holds if and only if $\forall p=1,2,...,P$, 
\begin{equation} \label{Theorem_estimable_iif}
{\mathscr{R}}\left( {{{\bf{V}}_{{{\boldsymbol{\theta}} ^{(p)}}}}{ \bf \Lambda }_{{{\boldsymbol{\theta}}^{(p)}}}^T}\right) \subseteq  {\mathscr{R}}\left( {{{\bf{V}}_{{{\boldsymbol{\tau }}^{(p)}}}}{\bf{\Lambda }}_{{{\boldsymbol{\tau }}^{(p)}}}^T} \right).
\end{equation}
\end{theorem}
\begin{IEEEproof}
Let's first examine the term in the sum in (\ref{Inverse_FIM_from_identities_Simplified}). 
Noticing by (\ref{J_tau_p_matrix}), (\ref{B_p_matrix}) and (\ref{J_theta_matrix}), we can obtain
\begin{align} \notag
& {{\bf{J}}_{{{\cal A}_p}}} - {{\bf{B}}_p}{\bf{J}}_{{{\boldsymbol{\tau }}^{(p)}}}^{ - 1}{\bf{B}}_p^T \\ \label{J_A_p_minus}
& \!\! = \! {{\bf{\Phi }}_{{{\boldsymbol{\theta }}^{(p)}}}}{\bf{\Phi }}_{{{\boldsymbol{\theta }}^{(p)}}}^T \!\! - \! {{\bf{\Phi }}_{{{\boldsymbol{\theta }}^{(p)}}}}{\bf{\Phi }}_{{{\boldsymbol{\tau }}^{(p)}}}^T{\left( {{{\bf{\Phi }}_{{{\boldsymbol{\tau }}^{(p)}}}}{\bf{\Phi }}_{{{\boldsymbol{\tau }}^{(p)}}}^T} \right)^{ - 1}}{{\bf{\Phi }}_{{{\boldsymbol{\tau }}^{(p)}}}}{\bf{\Phi }}_{{{\boldsymbol{\theta }}^{(p)}}}^T.
\end{align}
Denote
\begin{equation} \label{definition_D_Theorem2}
{\bf{D}} \buildrel \Delta \over = {\left( {{{\bf{\Phi }}_{{{\boldsymbol{\tau }}^{(p)}}}}{\bf{\Phi }}_{{{\boldsymbol{\tau }}^{(p)}}}^T} \right)^{ - 1}}{{\bf{\Phi }}_{{{\boldsymbol{\tau }}^{(p)}}}}{\bf{\Phi }}_{{{\boldsymbol{\theta }}^{(p)}}}^T,
\end{equation}
then by employing (\ref{J_A_p_minus}), we can obtain that
\begin{align} \notag
& {{\bf{J}}_{{{\cal A}_p}}} - {{\bf{B}}_p}{\bf{J}}_{{{\boldsymbol{\tau }}^{(p)}}}^{ - 1}{\bf{B}}_p^T \\ \notag
& = {\left( {{\bf{\Phi }}_{{{\boldsymbol{\theta }}^{(p)}}}^T - {\bf{\Phi }}_{{{\boldsymbol{\tau }}^{(p)}}}^T{\bf{D}}} \right)^T}\left( {{\bf{\Phi }}_{{{\boldsymbol{\theta }}^{(p)}}}^T - {\bf{\Phi }}_{{{\boldsymbol{\tau }}^{(p)}}}^T{\bf{D}}} \right) \\ \label{Proof_theorem_estimable_psd}
& \succeq {\bf{0}}.
\end{align}
What's more, the equality in (\ref{Proof_theorem_estimable_psd}) is attained if and only if
\begin{equation} \label{proof_theorem2_temp}
{ {{\bf{\Phi }}_{{{\boldsymbol{\theta }}^{(p)}}}^T - {\bf{\Phi }}_{{{\boldsymbol{\tau }}^{(p)}}}^T{\bf{D}}} } = {\bf{0}}, \forall p \ge 1.
\end{equation}
By employing (\ref{Phi_SVD}) and (\ref{definition_D_Theorem2}), (\ref{proof_theorem2_temp}) is equivalent to $\forall p \ge 1$,
\begin{equation}  \notag
{{\bf{V}}_{{{\boldsymbol{\tau }}^{(p)}}}}\!\!\left[ {{\bf{I}}\! -\! {\bf{\Lambda }}_{{{\boldsymbol{\tau }}^{(p)}}}^T{{\left( {{{\bf{\Lambda }}_{{{\boldsymbol{\tau }}^{(p)}}}}{\bf{\Lambda }}_{{{\boldsymbol{\tau }}^{(p)}}}^T} \right)}^{ - 1}}{{\bf{\Lambda }}_{{{\boldsymbol{\tau }}^{(p)}}}}} \right]\!{\bf{V}}_{{{\boldsymbol{\tau }}^{(p)}}}^T{{{\bf{V}}_{{{\boldsymbol{\theta}} ^{(p)}}}}{ \bf \Lambda }_{{{\boldsymbol{\theta}}^{(p)}}}^T} \!= \!{\bf{0}},
\end{equation}
which implies
\begin{equation} \label{Proof_theorem_estimable_equality_iif_condition}
{\mathscr{R}}\left( {{{\bf{V}}_{{{\boldsymbol{\theta}} ^{(p)}}}}{ \bf \Lambda }_{{{\boldsymbol{\theta}}^{(p)}}}^T} \right) \subseteq {\mathscr{R}}\left( {{{\bf{V}}_{{{\boldsymbol{\tau }}^{(p)}}}}{\bf{\Lambda }}_{{{\boldsymbol{\tau }}^{(p)}}}^T} \right), \forall p \ge 1.
\end{equation}
Consequently, from (\ref{Inverse_FIM_from_identities_Simplified}), (\ref{Proof_theorem_estimable_psd}), and (\ref{Proof_theorem_estimable_equality_iif_condition}), we can conclude that
\begin{equation} 
{\left[ {{\bf{J}}_{\boldsymbol{\Theta }}^{ - 1}} \right]_{1:{D_{\boldsymbol{\theta }}}}} \preceq {\bf{J}}_{{\cal A}_0}^{ - 1},
\end{equation}
with equality if and only if $\forall p=1,2,...,P$,
\begin{equation} 
{\mathscr{R}}\left( {{{\bf{V}}_{{{\boldsymbol{\theta}} ^{(p)}}}}{ \bf \Lambda }_{{{\boldsymbol{\theta}}^{(p)}}}^T} \right)  \subseteq {\mathscr{R}}\left( {{{\bf{V}}_{{{\boldsymbol{\tau }}^{(p)}}}}{\bf{\Lambda }}_{{{\boldsymbol{\tau }}^{(p)}}}^T} \right).
\end{equation}
\end{IEEEproof}

In \emph{Theorem \ref{Theorem_estimable_FIM_upper_bound}}, we provide the necessary and sufficient conditions under which the estimable spoofing attacks can deteriorate the CRB for estimating $\boldsymbol{\theta}$ to its upper bound as shown in (\ref{Upper_bound_J_theta_estimable}). We formally define this class of optimal estimable spoofing attacks next. 
\begin{definition}[Optimal Estimable Spoofing Attack] \label{Definition_Optimal_Estimable_Measurement_Attack}
An estimable spoofing attack which satisfies the necessary and sufficient condition in (\ref{Theorem_estimable_iif}) is called an optimal estimable spoofing attack (OESA).
\end{definition}

The physical meanings of the terms in (\ref{Inverse_FIM_from_identities_Simplified}) and the insight into \emph{Theorem \ref{Theorem_estimable_FIM_upper_bound}} deserve some discussion. The term ${{{\bf{J}}_{{{\cal A}_0}}}} $ represents the information on $\boldsymbol{\theta}$ embedded in the data from ${\cal A}_0$, while ${{\bf{J}}_{{{\cal A}_p}}}$ indicates the information on $\boldsymbol{\theta}$ that can be provided by the data from ${\cal A}_p$ if ${\boldsymbol{\tau}}^{(p)}$ is known to the FC. The term ${{\bf{B}}_p}{\bf{J}}_{{{\boldsymbol{\tau }}^{(p)}}}^{ - 1}{\bf{B}}_p^T$ specifies the degradation of the information on $\boldsymbol{\theta}$ from ${\cal A}_p$, which is induced by the uncertainty of ${\boldsymbol{\tau}}^{(p)}$. By considering the interpretations of these terms, the insight into \emph{Theorem \ref{Theorem_estimable_FIM_upper_bound}} is that if and only if (\ref{Theorem_estimable_iif}) holds, the uncertainty of ${\boldsymbol{\tau}}^{(p)}$ can reduce the information on $\boldsymbol{\theta}$ conveyed by the data from ${\cal A}_p$ to $0$ in which case the sum in the inverse does not contribute to (\ref{Inverse_FIM_from_identities_Simplified}). Moreover, \emph{Theorem \ref{Theorem_estimable_FIM_upper_bound}} points out that the degradation ${{\bf{B}}_p}{\bf{J}}_{{{\boldsymbol{\tau }}^{(p)}}}^{ - 1}{\bf{B}}_p^T$ cannot be strictly larger than ${{\bf{J}}_{{{\cal A}_p}}}$.

There is another interesting interpretation of (\ref{Theorem_estimable_iif}). We define the pmf vector ${\boldsymbol{\psi }}_p^{(j,k)}$ of the $k$-th measurement at the $j$-th sensor which is under the $p$-th spoofing attack as 
\begin{equation}
{\boldsymbol{\psi }}_p^{(j,k)} \buildrel \Delta \over = {\left[ { {p_{j1}^{(k)}} , {p_{j2}^{(k)}} ,..., {p_{j{R_j}}^{(k)}} } \right]^T},
\end{equation}
where the after-attack pmf $p_{jr}^{(k)}$ is defined in (\ref{pjr}).
It can be shown that (\ref{Theorem_estimable_iif}) is equivalent to the existence of a vector ${{\boldsymbol{\alpha }}^{(p,i)}} = {[ {\alpha _1^{(p,i)},\alpha _2^{(p,i)},...,\alpha _{{D_p}}^{(p,i)}} ]^T}$ such that for all $j \in {\cal A}_p$ and all $k$,
\begin{equation} \label{change_in_theta_pmf}
\frac{{\partial {\boldsymbol{\psi }}_p^{(j,k)}}}{{\partial {\theta _i}}} = \sum\limits_{l = 1}^{{D_p}} {\alpha _l^{(p,i)} \frac{{\partial {\boldsymbol{\psi }}_p^{(j,k)}}}{{\partial \tau _l^{(p)}}} }.
\end{equation}  
The relationship in (\ref{change_in_theta_pmf}) demonstrates that for all $j$ and all $k$, the change of the pmf vector ${\boldsymbol{\psi }}_p^{(j,k)}$ induced by changing each $\theta_i$ can be reproduced by a linear combination of the changes of the pmf vector ${\boldsymbol{\psi }}_p^{(j,k)}$ induced by changing the elements of ${\boldsymbol{\tau}}^{(p)}$.
This implies the FC will be unable to distinguish changes in the attack vector parameter $\boldsymbol{\tau}^{(p)}$ from changes in the desired vector parameter $\boldsymbol{\theta}$, based on the observations, which severely hinders estimation.

\emph{Theorem \ref{Theorem_estimable_FIM_upper_bound}} also describes how to design optimal estimable spoofing attacks. The adversaries choose $\{ {g_{jk}}( {x_{jk}| {{\boldsymbol{\theta }},{{\boldsymbol{\tau }}^{(p)}}} } )\}$ to meet the necessary and sufficient condition in (\ref{Theorem_estimable_iif}). One trivial example of OESA, which may be relatively easy to detect, is to replace the original measurements at the attacked sensors by some regenerated data obeying a distribution not parameterized by $\boldsymbol{\theta}$, which leads to  ${\boldsymbol{\Phi}}_{{\boldsymbol{\theta}}^{(p)}} = {\bf{0}}$ for all $p \ge 1$, and therefore, (\ref{Theorem_estimable_iif}) is satisfied. In the following part, some typical OESA examples of practical interest are investigated.

\begin{corollary} \label{Corollary_shift_parameter}
If the spoofing attacks are such that for any $p \ge 1$, $\exists {\lambda}_p $ satisfying
\begin{equation} \label{Corollary_condition}
 {{\bf{\Phi }}_{{{\boldsymbol{\theta }}^{(p)}}}} = {\lambda}_p  {{\bf{\Phi }}_{{{\boldsymbol{\tau }}^{(p)}}}},
\end{equation}
then the CRB ${{[ {{\bf{J}}_{\bf{\Theta }}^{ - 1}} ]}_{1:{D_{\boldsymbol{\theta }}}}}$ for $\boldsymbol{\theta}$ will be maximized in the positive semidefinite sense, more specifically
\begin{equation}
{\left[ {{\bf{J}}_{\boldsymbol{\Theta }}^{ - 1}} \right]_{1:{D_{\boldsymbol{\theta }}}}} = {\bf{J}}_{{{\cal A}_0}}^{ - 1}.
\end{equation}
Furthermore, 
the necessary and sufficient condition under which (\ref{Corollary_condition}) is satisfied for any ${\boldsymbol{\theta }}$, ${{\boldsymbol{\tau }}^{(p)}}$ and $\{I_j^{(r)}\}$ is that $\forall j \in {\cal A}_p$ and for all $k$, 
the after-attack pdf ${g_{jk}}( {x_{jk}| {{\boldsymbol{\theta }},{{\boldsymbol{\tau }}^{(p)}}} } )$ can be expressed as
\begin{equation} \label{Corollary_shift_parameter_gj}
{g_{jk}}\left( {x_{jk}\left| {{\boldsymbol{\theta }},{{\boldsymbol{\tau }}^{(p)}}} \right.} \right) = {{\tilde g}_{jk}}\left( {x_{jk}\left| {{\lambda}_p {\boldsymbol{\theta }} + {{\boldsymbol{\tau }}^{(p)}}} \right.} \right),
\end{equation}
for some ${{\tilde g}_{jk}}$.
\end{corollary}
\begin{IEEEproof}
Note that if for any $p \ge 1$, $\exists {\lambda _p} $ such that
\begin{equation}  \label{Proof_Corollary1_1}
{{\bf{\Phi }}_{{{\boldsymbol{\theta }}^{(p)}}}} = {\lambda _p} {{\bf{\Phi }}_{{{\boldsymbol{\tau }}^{(p)}}}},
\end{equation}
then $\forall p \ge 1$, $D_{\boldsymbol{\theta}} = D_p$ and
\begin{equation} \notag
{\mathscr{R}}\left( {{\bf{V}}_{{{\boldsymbol{\theta }}^{(p)}}}}{\bf{\Lambda }}_{{{\boldsymbol{\theta }}^{(p)}}}^T \right) \subseteq  {\mathscr{R}}\left( {{{\bf{V}}_{{{\boldsymbol{\tau }}^{(p)}}}}{\bf{\Lambda }}_{{{\boldsymbol{\tau }}^{(p)}}}^T} \right).
\end{equation}
Thus, by \emph{Theorem \ref{Theorem_estimable_FIM_upper_bound}}, we can obtain that
\begin{equation}
{\left[ {{\bf{J}}_{\boldsymbol{\Theta }}^{ - 1}} \right]_{1:{D_{\boldsymbol{\theta }}}}} = {\bf{J}}_{{\cal A}_0}^{ - 1}.
\end{equation}
In addition, by employing (\ref{Phi_theta_p}), (\ref{Phi_tau_p}) and (\ref{phi_theta_tau_p}), (\ref{Proof_Corollary1_1}) is equivalent to that for all $j \in {\cal A}_p$, all $k$, and all $r$,  
\begin{equation} \label{partial_p_theta}
\frac{{\partial p_{jr}^{(k)}}}{{\partial {\boldsymbol{\theta }}}} = {\lambda_p} \frac{{\partial p_{jr}^{(k)}}}{{\partial {{\boldsymbol{\tau }}^{(p)}}}}. 
\end{equation}
Noticing by (\ref{pjr}), in order to render (\ref{partial_p_theta}) be assured for any $\boldsymbol{\theta}$, ${\boldsymbol{\tau}}^{(p)}$ and $\{{I_j^{(r)}}\}$, the adversaries need to ensure that for any $\boldsymbol{\theta}$ and ${\boldsymbol{\tau}}^{(p)}$,
\begin{equation} \label{partial_g_theta_tau}
{\frac{\partial }{{\partial {\boldsymbol{\theta }}}}{g_{jk}}\left( {x_{jk}\left| {{\boldsymbol{\theta }},{{\boldsymbol{\tau }}^{(p)}}} \right.} \right)}  = {\lambda_p}  {\frac{\partial }{{\partial {{\boldsymbol{\tau }}^{(p)}}}}{g_{jk}}\left( {x_{jk}\left| {{\boldsymbol{\theta }},{{\boldsymbol{\tau }}^{(p)}}} \right.} \right)}
\end{equation}
for all $j \in {\cal A}_p$ and all $k$.

It is clear that if 
\begin{equation}
{g_{jk}}\left( {x_{jk}\left| {{\boldsymbol{\theta }},{{\boldsymbol{\tau }}^{(p)}}} \right.} \right) = {{\tilde g}_{jk}}\left( {x_{jk}\left| {{\lambda_p} {\boldsymbol{\theta }} + {{\boldsymbol{\tau }}^{(p)}}} \right.} \right),
\end{equation}
for some ${{\tilde g}_{jk}}$, then (\ref{partial_g_theta_tau}) holds. On the other hand, if (\ref{partial_g_theta_tau}) is true for any $\boldsymbol{\theta}$ and ${\boldsymbol{\tau}}^{(p)}$, then $\forall l=1,2,...,D_{\boldsymbol{\theta}}$,
\begin{align} \notag
& (1,  - {\lambda_p} )\left(  \begin{array}{l}
\frac{\partial }{{\partial {\theta _l}}}{g_{jk}}\left( {x_{jk} \left| {{{\{ {\theta _m}\} }_{m \ne l}},{{\{ \tau _m^{(p)}\} }_{m \ne l}},{\theta _l},\tau _l^{(p)}} \right.}  \right)\\
\frac{\partial }{{\partial \tau _l^{(p)}}}{g_{jk}}\left( {x_{jk}  \left| {{{\{ {\theta _m}\} }_{m \ne l}},{{\{ \tau _m^{(p)}\} }_{m \ne l}},{\theta _l},\tau _l^{(p)}} \right.}  \right)
\end{array}  \right) \\ \label{proof_corollary1_temp1}
& =  0
\end{align}
for any $\boldsymbol{\theta}$ and ${\boldsymbol{\tau}}^{(p)}$. This implies that for any ${\theta _l}$ and $\tau _l^{(p)}$, the gradient of ${g_{jk}}( {x_{jk}| {{{\{ {\theta _m}\} }_{m \ne l}},{{\{ \tau _m^{(p)}\} }_{m \ne l}},{\theta _l},\tau _l^{(p)}} } )$ with respect to $[{\theta _l}, \tau _l^{(p)}]^T$ is perpendicular to the vector $[1,  - {\lambda_p} ]^T$. Thus, if we change $[{\theta _l}, \tau _l^{(p)}]^T$ in the direction $[1,  - {\lambda_p} ]^T$, then ${g_{jk}}( {x_{jk}| {{{\{ {\theta _m}\} }_{m \ne l}},{{\{ \tau _m^{(p)}\} }_{m \ne l}},{\theta _l},\tau _l^{(p)}} } )$ does not change. Thus, any equivalent change in the perpendicular direction to $[1,  - {\lambda_p} ]^T$ will produce the same change in ${g_{jk}}( {x_{jk}| {{{\{ {\theta _m}\} }_{m \ne l}},{{\{ \tau _m^{(p)}\} }_{m \ne l}},{\theta _l},\tau _l^{(p)}} } )$. Therefore, for any $l$, if 
\begin{equation}
({\lambda_p} ,1)\left( \begin{array}{l}
0\\
t
\end{array} \right) = ({\lambda_p} ,1)\left( \begin{array}{l}
{\theta _l}\\
{\tau _l^{(p)}}
\end{array} \right),
\end{equation}
that is, $t = {\lambda_p} {\theta _l} + \tau _l^{(p)}$, then we can obtain that
\begin{align} \notag
& {g_{jk}}\left( {x_{jk}\left| {{{\{ {\theta _m}\} }_{m \ne l}},{{\{ \tau _m^{(p)}\} }_{m \ne l}},0,t} \right.} \right)\\ \label{proof_corollary_temp1}
&  = {g_{jk}}\left( {x_{jk}\left| {{{\{ {\theta _m}\} }_{m \ne l}},{{\{ \tau _m^{(p)}\} }_{m \ne l}},{\theta _l},\tau _l^{(p)}} \right.} \right).
\end{align}
As a result, for any $l$, by employing (\ref{proof_corollary_temp1}) and defining
\begin{align} \notag
& {{\bar g}_{jk,l}}\left( {x_{jk}\left| {{{\{ {\theta _m}\} }_{m \ne l}},{{\{ \tau _m^{(p)}\} }_{m \ne l}},t} \right.} \right) \\
& \buildrel \Delta \over = {g_{jk}}\left( {x_{jk}\left| {{{\{ {\theta _m}\} }_{m \ne l}},{{\{ \tau _m^{(p)}\} }_{m \ne l}},0,t} \right.} \right),
\end{align}
we can express ${g_{jk}}( {x_{jk}| {{{\{ {\theta _m}\} }_{m \ne l}},{{\{ \tau _m^{(p)}\} }_{m \ne l}},{\theta _l},\tau _l^{(p)}} } )$ as
\begin{align} \notag
& {g_{jk}}\left( {x_{jk}\left| {{{\{ {\theta _m}\} }_{m \ne l}},{{\{ \tau _m^{(p)}\} }_{m \ne l}},{\theta _l},\tau _l^{(p)}} \right.} \right) \\
& = {{\bar g}_{jk,l}}\left( {x_{jk}\left| {{{\{ {\theta _m}\} }_{m \ne l}},{{\{ \tau _m^{(p)}\} }_{m \ne l}}, {\lambda_p}  {\theta _l} + \tau _l^{(p)}} \right.} \right)
\end{align}
for some ${{\bar g}_{jk,l}}$, which implies that
\begin{equation}
{g_{jk}}\left( {x_{jk}\left| {{\boldsymbol{\theta }},{{\boldsymbol{\tau }}^{(p)}}} \right.} \right) = {{\tilde g}_{jk}}\left( {x_{jk}\left| {{\lambda_p} {\boldsymbol{\theta }} + {{\boldsymbol{\tau }}^{(p)}}} \right.} \right)
\end{equation}
for some ${{\tilde g}_{jk}}$.
\end{IEEEproof}

As demonstrated by \emph{Corollary \ref{Corollary_shift_parameter}}, if the spoofing attack gives rise to an after-attack pdf ${g_{jk}}( {x_{jk}| {{\boldsymbol{\theta }},{{\boldsymbol{\tau }}^{(p)}}} } )$ which is only parameterized by the sum of ${\lambda_p}\boldsymbol{\theta}$ and ${\boldsymbol{\tau}}^{(p)}$ for any $\lambda_p$, then the spoofing attack is optimal in the sense of \emph{Definition \ref{Definition_Optimal_Estimable_Measurement_Attack}}. This class of OESAs is interesting and powerful in practice, since their optimality is independent of the values of the desired vector parameter and the attack vector parameter. The DRFM example discussed in the introduction which introduces a time delay is one example of this class of OESAs (with $\lambda_p =1$).
For the scenario where the desired parameter is the mean of the observations, which is a popular signal model for sensor network estimation systems with quantized data \cite{ ribeiro2006bandwidth1, niu2006target, fang2009hyperplane, zhang2015Asymptotically}, this class of OESAs can be easily launched by just adding an offset to the measurements at each attacked sensor.


Another representative example of the class of OESAs described by (\ref{Corollary_shift_parameter_gj}) is extensively considered in smart grid systems under the name data-injection attacks, see \cite{liu2008attack, kim2011strategic, kosut2011malicious,  cui2012coordinated} and the references therein. At time instant $k$, the direct current power flow model in the absence of spoofing attacks can be expressed as
\begin{equation}
{{\bf{x}}_k} = {\bf{H}{\boldsymbol{\theta }}} + {{\bf{n}}_k}.
\end{equation}
Considering the $p$-th data-injection attack, the after-attack measurements from the sensors in ${\cal A}_p$ at time instant $k$ are given by
\begin{equation} \label{data_injection_attack}
{\left[ {{{{\bf{\tilde x}}}_k}} \right]_{{{\cal A}_p}}} = {\left[ {{{\bf{x}}_k}} \right]_{{{\cal A}_p}}} + {{\bf{a}}^{(p)}} = {\left[ {\bf{H}} \right]_{{{\cal A}_p},:}}{\boldsymbol{\theta }} + {{\bf{a}}^{(p)}} + {\left[ {{{\bf{n}}_k}} \right]_{{{\cal A}_p}}},
\end{equation}
where ${{{\bf{a}}^{(p)}}}$ represents the data injected by the $p$-th spoofing attack. 
If the adversaries choose ${{{\bf{a}}^{(p)}}}$ such that
\begin{equation}
{{\bf{a}}^{(p)}} = {\left[ {\bf{H}} \right]_{{{\cal A}_p},:}}{{\boldsymbol{\tau }}^{(p)}}
\end{equation}
for some ${{\boldsymbol{\tau }}^{(p)}}$, then the after-attack measurements from the sensors in ${\cal A}_p$ can be equivalently written as
\begin{equation}
{\left[ {{{{\bf{\tilde x}}}_k}} \right]_{{{\cal A}_p}}} = {\left[ {\bf{H}} \right]_{{{\cal A}_p},:}}\left( {{\boldsymbol{\theta }} + {{\boldsymbol{\tau }}^{(p)}}} \right) + {\left[ {{{\bf{n}}_k}} \right]_{{{\cal A}_p}}},
\end{equation}
and therefore, (\ref{Corollary_shift_parameter_gj}) is satisfied by the data-injection attack. Further, by \emph{Corollary \ref{Corollary_shift_parameter}}, the CRB for $\boldsymbol{\theta}$ is maximized in the positive semidefinite sense if all the attacks are of this type. Moreover, it can be shown that the stealth attack or undetectable attack in \cite{liu2008attack, kosut2011malicious, cui2012coordinated}, which attracts extensive attention in recent literature on smart grids, is just such an attack with $P=1$.

In addition to the class of OESAs described in (\ref{Corollary_shift_parameter_gj}), there are many other OESAs. For example, if the $p$-th spoofing attack satisfies that $\forall j \in {\cal A}_p$ and for all $k$, ${g_{jk}}( {x_{jk}| {{\boldsymbol{\theta }},{{\boldsymbol{\tau }}^{(p)}}} } ) = {{\tilde g}_{jk}}( {x_{jk}| {h_{jk} ( {{\boldsymbol{\theta }},{{\boldsymbol{\tau }}^{(p)}}} )} } )$ for some ${{\tilde g}_{jk}}$ and some symmetric function ${h_{jk}}$ of ${\boldsymbol{\theta }}$ and ${{\boldsymbol{\tau }}^{(p)}}$, then it can be shown that the $p$-th spoofing attack is an OESA provided that the values of ${{\boldsymbol{\tau }}^{(p)}}$ and ${\boldsymbol{\theta }}$ are equal.

\subsection{Discussion}

Under the conditions of \emph{Definition \ref{Definition_Opt_Measurement_Attacks}}, it is clear that ${\bf{J}}_{{{\cal A}_0}}^{ - 1}$ is an upper bound on the CRB for $\boldsymbol{\theta}$, no matter what kind of attacks have been launched. From 
(\ref{CRB_inestimable}) and \emph{Theorem \ref{Theorem_estimable_FIM_upper_bound}}, 
the CRB for $\boldsymbol{\theta}$ under ISA or OESA equals to its upper bound ${\bf{J}}_{{{\cal A}_0}}^{ - 1}$. Therefore, according to 
\emph{Definition \ref{Definition_Opt_Measurement_Attacks}}, both ISA and OESA are OGDSAs. Furthermore, note that ${{\bf \Lambda}_{{\boldsymbol{\tau}}^{(p)}}}$ is a $ D_p \times (\sum\nolimits_{j \in {{\cal A}_p}} {{R_j}})$ matrix, and hence, $\text{rank}({{\bf \Lambda}_{{\boldsymbol{\tau}}^{(p)}}}) \le D_p$. Thus, any OGDSA is either an ISA when $\text{rank}({{\bf \Lambda}_{{\boldsymbol{\tau}}^{(p)}}}) < D_p$, or an OESA when $\text{rank}({{\bf \Lambda}_{{\boldsymbol{\tau}}^{(p)}}}) = D_p$. 

A particular note of interest is that 
the results in \emph{Section \ref{Section_Inestimable_Spoofing_Attacks}} and \emph{\ref{Section_Optimal_Estimable_Spoofing_Attacks}} can be used to judge whether the attacked measurements are useful or not in terms of reducing CRB under the conditions of \emph{Definition \ref{Definition_Opt_Measurement_Attacks}}. In particular, it is seen from 
(\ref{CRB_inestimable}) and \emph{Theorem \ref{Theorem_estimable_FIM_upper_bound}} that the CRB for $\boldsymbol{\theta}$ in the presence of ISA or OESA is the same as the CRB for $\boldsymbol{\theta}$ when only unattacked data is used. Thus, we obtain the following corollary.
\begin{corollary}
Under the conditions of \emph{Definition \ref{Definition_Opt_Measurement_Attacks}}, the necessary and sufficient condition under which the attacked measurements are useless in terms of reducing CRB is that the spoofing attacks belong to either ISA or OESA which are defined in \emph{Definition \ref{Definition_inestimable_attack}} and \emph{\ref{Definition_Optimal_Estimable_Measurement_Attack}} respectively. 
\end{corollary}

However, the fundamental mechanisms of ISA and OESA for making the attacked measurements useless in terms of reducing CRB are very different. To be specific, ISA renders the task of estimating the attack vector parameters beyond the capabilities of the quantized estimation system by causing the FIM for jointly estimating the desired and attack vector parameters to be singular. Thus, ISA prevents the FC from potentially improving the CRB for $\boldsymbol{\theta}$ by jointly estimating $\boldsymbol{\theta}$ and the attack vector parameters. In contrast, 
even though the joint FIM is nonsingular,
the FC is not able to obtain any improvement from using the attacked data in the CRB performance for $\boldsymbol{\theta}$ under OESA. 

It is worth mentioning that 
(\ref{CRB_inestimable}) and \emph{Theorem \ref{Theorem_estimable_FIM_upper_bound}} demonstrate that under the conditions of \emph{Definition \ref{Definition_Opt_Measurement_Attacks}}, the CRB for $\boldsymbol{\theta}$ reaches its upper bound in the presence of ISA or OESA.
In practice, however, the FC may not be able to well identify the set of unattacked sensors and categorize the attacked sensors into different groups according to distinct types of spoofing attacks. Thus, the actual estimation performance under ISA and OESA can be expected to be inferior to ${\bf J}_{{\cal{A}}_0}^{-1}$.

In the case where for any $j$, $\{{\tilde{x}}_{jk}\}$ is a statistically independent and identically distributed sequence over $k$,
then in \emph{Theorem \ref{Theorem_inestimable_attack}}, the sufficient condition in (\ref{Theorem_inestimable_attack_sufficient_condition}) will become $D_p >  \sum\nolimits_{j \in {{\cal A}_p}} {{R_j}} - |{\cal A}_p|$. Now the right-hand side of the inequality is just the sum of the sizes of the alphabet sets employed at the sensors under the $p$-th spoofing attack minus the size of ${\cal A}_p$. This quantity even does not depend on the number of temporal measurements at each attacked sensor. In addition, it can be shown that the results in this section can be easily extended to the cases where the attack vector parameters employed at the attacked sensors change over time, that is, the after-attack observation ${\tilde{x}}_{jk}$ obeys the statistical model
\begin{equation}
{{\tilde x}_{jk}} \sim \left\{ \begin{array}{l}
{f_{jk}}\left( {{\tilde x_{jk}}\left| {\boldsymbol{\theta }} \right.} \right), \quad {\text{if }} j \in {{{\cal U}}} \\
{g_{jk}}\left( {{\tilde x_{jk}}\left| {{\boldsymbol{\theta }},{{\boldsymbol{\xi  }}^{(j,k)}}} \right.} \right), \; {\text{if }} j \in {{\cal V}}
\end{array} \right.,
\end{equation}
where ${{\boldsymbol{\xi  }}^{(j,k)}}$ is the attack vector parameter employed at time instant $k$ at the $j$-th sensor. For the sake of brevity, we skip the extension.

\section{Joint Attack Identification and Parameter Estimation Under Estimable Spoofing Attack}
\label{Sec_JointIdentification}

In order to corroborate the theory just described, we develop a representative approach for the joint identification of attacked sensors along with the estimation of the desired vector parameter and the attack vector parameters for the estimation system facing attacks.
In this section, we focus on a class of estimable spoofing attacks
in which for any $p$, $\forall j \in {\cal A}_p$, the FIM for ${\boldsymbol{\tau}}^{(p)}$ based on the data from the $j$-th sensor is nonsingular. 
Further, we  assume that ${\bf{J}}_{{{\cal A}_0}}$ defined in (\ref{J_A_p}) is nonsingular in the presence of spoofing attacks. This could occur, for example, if only a small subset of sensors can be attacked in a distributed sensor setting or if a subset of sensors can be well protected in advance to give rise to a nonsingular ${\bf{J}}_{{{\cal A}_0}}$. 

In this section, we use $\{ {{{\boldsymbol{\xi  }}^{(j)}}} \}_{j = 1}^N$ instead of $\{ {{{\boldsymbol{\tau }}^{(p)}}} \}_{p = 1}^P$ to denote the attack vector parameters employed by the adversaries at the $j$-th sensor. 
For the sake of notational simplicity, we let $q_{jr}^{(k)}$ and ${{\tilde q}_{jr}}^{(k)}$ to denote the $r$-th value of the after-attack pmf of the $k$-th time sample at the $j$-th sensor when it is unattacked and attacked respectively 
\begin{equation} \label{definition_q_jr_unattack}
	{q_{jr}^{(k)}}  \buildrel \Delta \over =  \int_{I_j^{(r)}}  {{f_{jk}}\left( {x_{jk}\left| {\boldsymbol{\theta }} \right.} \right)dx_{jk}}
\end{equation}
and
\begin{equation} \label{definition_q_jr_attack}
 {{\tilde q}_{jr}^{(k)}}   \buildrel \Delta \over =   \int_{I_j^{(r)}}  {{{ g}_{jk}}\left( {x_{jk}\left| {{\boldsymbol{\theta }}, {{\boldsymbol{\xi  }}^{(j)}}} \right.} \right)dx_{jk}},
\end{equation}
where ${f_{jk}}$ and ${{ g}_{jk}}$ represent the pdf of the $k$-th time sample at the $j$-th sensor when it is unattacked and attacked respectively. 

Before proceeding, the following assumptions are made from a practical viewpoint in this section.
\begin{assumption} \label{Assumption_More_than_half}
As the sensors are assumed to be spread over a wide area and typically adversaries have limited resources, we assume that no more than half of sensors are attacked. 
\end{assumption}
\begin{assumption}[Significant Attack] \label{Assumption_Significant_Attack}
In order to give rise to sufficient impact on the statistical characterization of the measurements at each attacked sensors, every attacker is required to guarantee a minimum average distortion of the pmf at each attacked sensor, that is, 
\begin{equation} \label{impact_assumption_pmf}
\frac{1}{{{K_j}}}\sum\limits_{k = 1}^{{K_j}} {{{\left\| {{\bf{\tilde q}}_j^{(k)} - {\bf{q}}_j^{(k)}} \right\|}_2}}  \ge {d_q}, \forall j \notin {\cal A}_0,
\end{equation}
where ${{\bf{ q}}_j^{(k)}}$ and ${{\bf{\tilde q}}_j^{(k)}}$ are defined as
\begin{equation}
{\bf{q}}_j^{(k)} \buildrel \Delta \over = {\left[ {q_{j1}^{(k)},q_{j2}^{(k)},...,q_{j{R_j}}^{(k)}} \right]^T}
\end{equation}
and
\begin{equation}
{\bf{\tilde q}}_j^{(k)} \buildrel \Delta \over = {\left[ {\tilde q_{j1}^{(k)},\tilde q_{j2}^{(k)},...,\tilde q_{j{R_j}}^{(k)}} \right]^T}.
\end{equation}
\end{assumption}
We do not consider modifications smaller than  (\ref{impact_assumption_pmf}) as attacks and assume they have little impact on performance. \emph{Assumption \ref{Assumption_More_than_half}} is made for avoiding the ambiguity between the attacked and unattacked sensors.

Let ${\bf{\Omega }}$ denote a vector containing the desired vector parameter $\boldsymbol{\theta}$, the set of unknown attack vector parameters $\{{\boldsymbol{\xi}}^{(j)}\}$ as well as a set of unknown binary state variables $\{{\eta}_j\}$ that
\begin{equation}
{\bf{\Omega }} \buildrel \Delta \over = {\left[ {{\bf{\Xi }}^T,{{\boldsymbol{\eta }}^T}} \right]^T},
\end{equation}
where
\begin{equation}
{{\bf{\Xi }} } \buildrel \Delta \over = {\left[ {{{\boldsymbol{\theta }}^T},{{\left( {{{\boldsymbol{\xi }}^{(1)}}} \right)}^T},{{\left( {{{\boldsymbol{\xi }}^{(2)}}} \right)}^T},...,{{\left( {{{\boldsymbol{\xi }}^{(N)}}} \right)}^T}} \right]^T}
\end{equation}
and 
\begin{equation}
{\boldsymbol{\eta }} \buildrel \Delta \over = {\left[ {{{\eta}_1},{{\eta}_2},...,{{\eta}_N}} \right]^T}.
\end{equation}
The $j$-th element of ${\boldsymbol{\eta }}$ is zero, i.e., ${\eta}_j=0$, if the $j$-th sensor is unattacked, while ${\eta}_j=1$ implies the $j$-th sensor is attacked. The log-likelihood function evaluated at ${{\bf{\tilde u}}} = {\bf{r}}$ is
\begin{align} \notag
L\left( {\bf{\Omega }} \right) & \buildrel \Delta \over = \ln \Pr \left( {{\bf{\tilde u}} = {\bf{r}}\left| {\bf{\Omega }} \right.} \right)\\ \label{loglikelihood_Omega} 
&  = \sum\limits_{j = 1}^N {\sum\limits_{k = 1}^{{K_j}} {\left[ {{\eta_j}\ln {{\tilde q}_{jr_{jk}}^{(k)}} + \left( {1 - {\eta_j}} \right)\ln {q_{j{r_{jk}}}^{(k)}}} \right]} }. 
\end{align}
Based on this setting, the FC can jointly identify the state of each sensor and estimate the desired vector parameter $\boldsymbol{\theta}$ by solving the following constrained optimization problem
\begin{subequations} \label{Optimization_Identification_Estimation}
\begin{align} \label{Optimization_Identification_Estimation_Object}
{\bf{\hat \Omega }} = \arg   \mathop {\max } \limits_{\bf{\Omega }} & \sum\limits_{j = 1}^N {\sum\limits_{k = 1}^{{K_j}} {\left[ {{{\eta}_j}\ln {{\tilde q}_{j{r_{jk}}}^{(k)}} + \left( {1 - {{\eta}_j}} \right)\ln {q_{j{r_{jk}}}^{(k)}}} \right]} } \\ \label{Optimization_Constraint_Ij}
 \text{s. t. } \; &  {{\eta}_j} \in \left\{ {0,1} \right\}, \; \forall j, \\ \label{Optimization_Constraint_Sum_Ij}
 & \sum\limits_{j = 1}^N {{{\eta}_j}}  < \frac{N}{2}, \\ \label{Optimization_Constraint_tau_j}
& \frac{1}{{{K_j}}}\sum\limits_{k = 1}^{{K_j}} {{{\left\| {{\bf{\tilde q}}_j^{(k)} - {\bf{q}}_j^{(k)}} \right\|}_2}}  \ge {d_q}, \; \forall {{\eta}_j} = 1,
\end{align}
\end{subequations}
where the constraints in (\ref{Optimization_Constraint_Sum_Ij}) and (\ref{Optimization_Constraint_tau_j}) are due to \emph{Assumption \ref{Assumption_More_than_half}} and \emph{Assumption \ref{Assumption_Significant_Attack}}.  

The integer constraint in (\ref{Optimization_Constraint_Ij}) makes the optimization problem difficult to solve. For small $N$, it may be solved exactly simply by exhaustively searching through all possible combinations of $\{{\eta}_j\}$, while for large $N$, this is not feasible in practice, since the number of all possible combination of $\{{\eta}_j\}$ is on the order of $2^N$. 
To this end, 
it is of considerable practical interest to develop an efficient algorithm
to solve the optimization problem in (\ref{Optimization_Identification_Estimation}). In this section, we propose a heuristic for solving (\ref{Optimization_Identification_Estimation}). 

\subsection{Random Relaxation with the EM Algorithm}
\label{Sec_Random_Relaxation}

According to the constraint in (\ref{Optimization_Constraint_Ij}), ${\eta}_j$ is an unknown deterministic binary variable, and hence, (\ref{Optimization_Constraint_Ij}) is equivalent to
\begin{equation} \label{Definition_pi_j}
{\pi_j} \buildrel \Delta \over = \Pr \left( {{{\eta}_j} \! = \! 1} \right) \in \{0,1\} {\text{ and }} \Pr \left( {{{\eta}_j} \! = \! 0} \right) = 1 - {\pi_j}, \forall j.
\end{equation}
Further, by dropping the constraint (\ref{Optimization_Constraint_Sum_Ij}) as well as (\ref{Optimization_Constraint_tau_j}), and then relaxing the deterministic $\{{\eta}_j\}$ to be random, that is, allowing $\pi_j = \Pr \left( {{{\eta}_j} = 1} \right) \in [0,1]$ for all $j=1,2,...,N$, the problem in (\ref{Optimization_Identification_Estimation}) reduces to 
\begin{subequations} \label{Optimization_Identification_Estimation_1}
\begin{align} \label{Optimization_Identification_Estimation_Object_1}
{{{\bf{\hat \Omega }}}_{{\pi }}} =  \arg \mathop {\max } \limits_{{{\bf{\Omega }}_{{\pi }}}} & \sum\limits_{j = 1}^N {\sum\limits_{k = 1}^{{K_j}} {\ln \left[ {{\pi _j}{{\tilde q}_{j{r_{jk}}}^{(k)}} + \left( {1 - {\pi _j}} \right){q_{j{r_{jk}}}^{(k)}}} \right]} }  \\ \label{Optimization_Constraint_Ij_1}
 \text{s. t. } \; &  {\pi_j} \in  [0,1], \; \forall j=1,2,...,N,
\end{align}
\end{subequations}
where ${{\bf{\Omega }}_\pi } \buildrel \Delta \over = {[ {{\bf{\Xi }}^T,{{\boldsymbol{\pi }}^T}} ]^T}$ and ${\boldsymbol{\pi }} \buildrel \Delta \over =  {\left[ {{\pi _1},{\pi _2}...,{\pi _N}} \right]^T}$.

The physical interpretation behind (\ref{Optimization_Identification_Estimation_1}) is that via random relaxation of the deterministic binary vector state variable $\boldsymbol{\eta}$, the set ${{\cal A}_0}$ of unattacked sensors is no longer deterministic, and moreover, each sensor in the sensor network is attacked with a certain probability $\pi_j$ at every time instant. 

By introducing a latent vector variable
\begin{equation}
{\bf{z}} = {\left[ {{z_{11}},{z_{12}},...,{z_{1{K_1}}},{z_{21}},...,{z_{N{K_N}}}} \right]^T},
\end{equation}
where $z_{jk}=1$ indicates that the $k$-th measurement at the $j$-th sensor was attacked, and $z_{jk}=0$ implies that the $k$-th measurement at the $j$-th sensor was unattacked, we can employ the Expectation-Maximization (EM) algorithm \cite{dempster1977maximum, mclachlan1997algorithm}, which is an iterative method that alternates between performing an expectation (E) step and a maximization (M) step, to solve the relaxed problem in (\ref{Optimization_Identification_Estimation_1}).

\subsubsection{E-step}

The \emph{E-step} computes the expected log-likelihood function $Q( {{{\bf{\Omega }}_\pi }| {{\bf{\Omega }}'_\pi } })$, with respect to $\bf{z}$ given the quantized data ${{\bf{\tilde u}} = {\bf{r}}}$ and the current estimate of the vector parameter ${\bf{\hat \Omega }}'_\pi  = {[ {{{( {{\bf{\hat \Xi '}}} )^T}},{{( {{\boldsymbol{\hat \pi }'}} )^T}}} ]^T}$, as following
\begin{equation} \label{Definition_Q}
Q\left( {{{\bf{\Omega }}_\pi }\left| {{{\bf{\hat \Omega }}}'_\pi } \right.} \right) \buildrel \Delta \over = {{\mathbbm{E}}_{\left. {\bf{z}} \right|{\bf{ \hat \Omega }}'_\pi,{\bf{\tilde u}} = {\bf{r}}}}\left\{ {L\left( {{{\bf{\Omega }}_\pi }} \right)} \right\},
\end{equation}
where the log-likelihood function ${L\left( {{{\bf{\Omega }}_\pi }} \right)}$  is given by
\begin{align} \notag
 L\left( {{{\bf{\Omega }}_\pi }} \right) & = \ln \Pr \left( {\left. {{\bf{z}},{\bf{\tilde u}} = {\bf{r}}} \right|{{\bf{\Omega }}_\pi }} \right)\\ \notag
& = \ln \Pr \left( {\left. {{\bf{\tilde u}} = {\bf{r}}} \right|{{\bf{\Omega }}_\pi },{\bf{z}}} \right) + \ln \Pr \left( {\left. {\bf{z}} \right|{{\bf{\Omega }}_\pi }} \right)\\  \notag
&  = \sum\limits_{j = 1}^N {\sum\limits_{k = 1}^{{K_j}} {\left\{ {{{\mathbbm{1}}_{\left\{ {{z_{jk}} = 1} \right\}}}\left( {\ln {{\tilde q}_{j{r_{jk}}}^{(k)}} + \ln {\pi _j}} \right)} \right.} } \\ \label{E_step_LLF} 
& \qquad  \quad \left. { + {{\mathbbm{1}}_{\left\{ {{z_{jk}} = 0} \right\}}}\left[ {\ln {q_{j{r_{jk}}}^{(k)}} + \ln \left( {1 - {\pi _j}} \right)} \right]} \right\}.
\end{align}
Define
\begin{align} \label{Definition_upsilon_1}
	\upsilon _{jk}^{(1)}  \!  \buildrel \Delta \over =  \!   {{\mathbbm{E}}_{\left. {\bf{z}} \right|{\bf{\hat \Omega }}'_\pi ,{\bf{\tilde u}} = {\bf{r}}}}\left\{ {{{\mathbbm{1}}_{\left\{ {{z_{jk}}  \!  =  \!  1} \right\}}}} \right\}  \!  = \!   \frac{{{{\hat{\pi}}' _j}{{\tilde q}_{j{r_{jk}}}^{(k)}}}}{{{{\hat{\pi}}' _j}{{\tilde q}_{j{r_{jk}}}^{(k)}}  \!   \!  +  \!     \left( {1 \! - {{\hat{\pi}}' _j}} \right){q_{j{r_{jk}}}^{(k)}}}}
\end{align}
and
\begin{equation} \label{Definition_upsilon_0}
\upsilon _{jk}^{(0)} \buildrel \Delta \over = {{\mathbbm{E}}_{\left. {\bf{z}} \right|{\bf{\hat \Omega }}'_\pi,{\bf{\tilde u}} = {\bf{r}}}}\left\{ {{{\mathbbm{1}}_{\left\{ {{z_{jk}} = 0} \right\}}}} \right\} = 1 - \upsilon _{jk}^{(1)},
\end{equation}
then by employing (\ref{Definition_Q}) and (\ref{E_step_LLF}), we can obtain the expected log-likelihood function
\begin{align} \notag
Q\left( {{{\bf{\Omega }}_\pi }\left| {{\bf{\hat \Omega }}'_\pi} \right.} \right) & = \sum\limits_{j = 1}^N {\sum\limits_{k = 1}^{{K_j}} {\left\{ {\upsilon _{jk}^{(1)}\left( {\ln {{\tilde q}_{j{r_{jk}}}^{(k)}} + \ln {\pi _j}} \right)} \right.} } \\
&  \quad \quad  \quad  \left. { + \upsilon _{jk}^{(0)}\left[ {\ln {q_{j{r_{jk}}}^{(k)}} + \ln \left( {1 - {\pi _j}} \right)} \right]} \right\}.
\end{align}

\subsubsection{M-step}

The \emph{M-step} seeks to find a new estimate of the vector parameter ${\bf{\hat \Omega}}_{\pi}$ to update the current estimate of the vector parameter ${\bf{\hat \Omega}}'_{\pi}$ by maximizing the expected log-likelihood function $Q( {{{\bf{\Omega }}_\pi }| {{\bf{\hat \Omega }}'_\pi } })$, that is,
\begin{equation} \label{M_step_update}
{{{\bf{\hat \Omega }}}_\pi } = {\left[ {{\bf{\hat \Xi }}_{}^T,{{{\boldsymbol{\hat \pi }}}^T}} \right]^T} = \arg \mathop {\max } \limits_{{{\bf{\Omega }}_{{\pi }}}}  Q\left( {{{\bf{\Omega }}_\pi }\left| {{\bf{\hat \Omega }}'_\pi} \right.} \right).
\end{equation}

\paragraph{Updated estimate of $\boldsymbol{\pi}$} 
According to (\ref{M_step_update}), the updated estimate ${\hat \pi}_j$ should satisfy
\begin{equation}
\frac{{\partial Q\left( {{{\bf{\Omega }}_\pi }\left| {{\bf{\hat \Omega }}'_\pi} \right.} \right)}}{{\partial {\pi _j}}} = \frac{1}{{{\pi _j}}}\sum\limits_{k = 1}^{{K_j}} {\upsilon _{jk}^{(1)}}  - \frac{1}{{1 - {\pi _j}}}\sum\limits_{k = 1}^{{K_j}} {\upsilon _{jk}^{(0)}}  = 0,
\end{equation}
which yields, by employing (\ref{Definition_upsilon_0}),
\begin{equation}
{{\hat \pi }_j} = \frac{1}{{{K_j}}}\sum\limits_{k = 1}^{{K_j}} {\upsilon _{jk}^{(1)}}.
\end{equation}

\paragraph{Updated estimate of $\boldsymbol{\Xi}$} 
\label{Section_Updated_Xi}

Similarly, the updated estimate $\boldsymbol{\hat \Xi}$ is the solution of the following equation
\begin{equation} \label{Update_Xi_condition}
{\nabla _{\bf{\Xi }}}Q\left( {{{\bf{\Omega }}_\pi }\left| {{\bf{\hat \Omega }}'_\pi} \right.} \right) = {\bf{0}}.
\end{equation}
Generally, a closed-form solution for the above equation may not exist. To solve (\ref{Update_Xi_condition}) in such cases, Newton's method can be employed with an initial point ${\bf{\hat \Xi }}^{(0)} = {{\bf{\hat \Xi }}'}$. At the $(i+1)$-th iteration of Newton's Method, the updated point ${\bf{\hat \Xi }}^{(t+1)}$ can be expressed as
\begin{align}  \notag
& {{{\bf{\hat \Xi }}}^{(t + 1)}} \\ \label{Newton_update}
& = {{{\bf{\hat \Xi }}}^{(t)}} - {\kappa}_t {\left[ {\nabla _{\bf{\Xi }}^2Q\left( {{\bf{\Omega }}_\pi ^{(t)}\left| {{\bf{\hat \Omega }}'_\pi} \right.} \right)} \right]^{ - 1}}{\nabla _{\bf{\Xi }}}Q\left( {{\bf{\Omega }}_\pi ^{(t)}\left| {{\bf{\hat \Omega }}'_\pi} \right.} \right)
\end{align}
where ${\bf{\Omega }}_\pi ^{(t)} = {[ {{{( {{{{\bf{\hat \Xi }}}^{(t)}}} )^T}},{{( {{\boldsymbol{\hat \pi '}}} )^T}}} ]^T}$,
and ${\kappa}_t \in (0,1)$ is the $t$-th step size computed by using a backtracking line search \cite{boyd2004convex}.

For completeness, the explicit expressions for the gradient and Hessian of the expected log-likelihood function with respect to ${\bf{\Xi }}$ are provided. The gradient ${\nabla _{\bf{\Xi }}}Q( {{{\bf{\Omega }}_\pi ^{(t)} }| {{\bf{\hat \Omega }}'_\pi} } )$ consists of the quantities $\frac{\partial }{{\partial {\theta _l}}}Q( {{\bf{\Omega }}_\pi ^{(t)}| {{\bf{\hat \Omega }}'_\pi} } )$ and $\frac{\partial }{{\partial \xi _l^{(j)}}}Q( {{\bf{\Omega }}_\pi ^{(t)}| {{\bf{\hat \Omega }}'_\pi} } ) $ for different $j$ and $l$, which can be computed by
\begin{align} \notag
& \frac{\partial }{{\partial {\theta _l}}}Q\left( {{\bf{\Omega }}_\pi ^{(t)}\left| {{\bf{\hat \Omega }}'_\pi} \right.} \right) \\ \label{partial_Q_partial_theta}
& = \sum\limits_{j = 1}^N {\sum\limits_{k = 1}^{{K_j}} {\left\{ {\upsilon _{jk}^{(1)}\frac{1}{{{{\tilde q}_{j{r_{jk}}}^{(k)}}}}\frac{\partial }{{\partial {\theta _l}}}{{\tilde q}_{j{r_{jk}}}^{(k)}} \! + \upsilon _{jk}^{(0)}\frac{1}{{{q_{j{r_{jk}}}^{(k)}}}}\frac{\partial }{{\partial {\theta _l}}}{q_{j{r_{jk}}}^{(k)}}} \right\}} } 
\end{align} 
and
\begin{align}
\frac{\partial }{{\partial \xi _l^{(j)}}}Q\left( {{\bf{\Omega }}_\pi ^{(t)}\left| {{\bf{\hat \Omega }}'_\pi} \right.} \right) = \sum\limits_{k = 1}^{{K_j}} {\upsilon _{jk}^{(1)}\frac{1}{{{{\tilde q}_{j{r_{jk}}}^{(k)}}}}\frac{\partial }{{\partial \xi _l^{(j)}}}{{\tilde q}_{j{r_{jk}}}^{(k)}}}.
\end{align}
The elements of the Hessian ${\nabla _{\bf{\Xi }}^2Q( {{\bf{\Omega }}_\pi ^{(t)}| {{\bf{\hat \Omega }}'_\pi} } )}$ can be calculated by the following expressions
\begin{align} \notag
& \frac{{{\partial ^2}}}{{\partial {\theta _l}\partial {\theta _m}}}Q\left( {{\bf{\Omega }}_\pi ^{(t)}\left| {{\bf{\hat \Omega }}'_\pi} \right.} \right) \\ \notag
& = \sum\limits_{j = 1}^N {\sum\limits_{k = 1}^{{K_j}} {\left\{ {\upsilon _{jk}^{(1)}\left( {\frac{1}{{\tilde q_{j{r_{jk}}}^{(k)}}}\frac{{{\partial ^2}{{\tilde q}_{j{r_{jk}}}^{(k)}}}}{{\partial {\theta _l}\partial {\theta _m}}} \! - \frac{1}{({\tilde q_{j{r_{jk}}}^{(k)}})^2}\frac{{\partial {{\tilde q}_{j{r_{jk}}}^{(k)}}}}{{\partial {\theta _l}}}\frac{{\partial {{\tilde q}_{j{r_{jk}}}^{(k)}}}}{{\partial {\theta _m}}}} \right)} \right.} } \\
& \;  \left. { + \upsilon _{jk}^{(0)}\left( {\frac{1}{{{q_{j{r_{jk}}}^{(k)}}}}\frac{{{\partial ^2}{q_{j{r_{jk}}}^{(k)}}}}{{\partial {\theta _l}\partial {\theta _m}}}  -  \frac{1}{({q_{j{r_{jk}}}^{(k)}})^2}\frac{{\partial {q_{j{r_{jk}}}^{(k)}}}}{{\partial {\theta _l}}}\frac{{\partial {q_{j{r_{jk}}}^{(k)}}}}{{\partial {\theta _m}}}} \right)} \right\},
\end{align}
\begin{align} \notag
& \frac{{{\partial ^2}}}{{\partial {\theta _l}\partial \xi _m^{(j)}}}Q\left( {{\bf{\Omega }}_\pi ^{(t)}\left| {{\bf{\hat \Omega }}'_\pi} \right.} \right) \\
& = \! \sum\limits_{k = 1}^{{K_j}} {\upsilon _{jk}^{(1)} \!  \left( {\frac{1}{{\tilde q_{j{r_{jk}}}^{(k)}}}\frac{{{\partial ^2}{{\tilde q}_{j{r_{jk}}}^{(k)}}}}{{\partial {\theta _l}\partial \xi _m^{(j)}}} \! - \! \frac{1}{({\tilde q_{j{r_{jk}}}^{(k)}})^2}\frac{{\partial {{\tilde q}_{j{r_{jk}}}^{(k)}}}}{{\partial {\theta _l}}}\frac{{\partial {{\tilde q}_{j{r_{jk}}}^{(k)}}}}{{\partial \xi _m^{(j)}}}} \right)}, 
\end{align}
\begin{align} \notag
& \frac{{{\partial ^2}}}{{\partial {\xi _l^{(j)}}\partial \xi _m^{(j)}}}Q\left( {{\bf{\Omega }}_\pi ^{(t)}\left| {{\bf{\hat \Omega }}'_\pi} \right.} \right) \\
& = \! \sum\limits_{k = 1}^{{K_j}} \! {\upsilon _{jk}^{(1)} \! \! \left( \! {\frac{1}{{\tilde q_{j{r_{jk}}}^{(k)}}}\frac{{{\partial ^2}{{\tilde q}_{j{r_{jk}}}^{(k)}}}}{{\partial {\xi _l^{(j)}}\partial \xi _m^{(j)}}} \! - \! \frac{1}{({\tilde q_{j{r_{jk}}}^{(k)}})^2}\frac{{\partial {{\tilde q}_{j{r_{jk}}}^{(k)}}}}{{\partial {\xi _l^{(j)}}}}\frac{{\partial {{\tilde q}_{j{r_{jk}}}^{(k)}}}}{{\partial \xi _m^{(j)}}}} \! \right)}, 
\end{align}
and
\begin{equation} \label{partial2_Q_partial_tau_i_partial_tau_j}
\frac{{{\partial ^2}}}{{\partial \xi _l^{(i)}\partial \xi _m^{(j)}}}Q\left( {{\bf{\Omega }}_\pi ^{(t)}\left| {{\bf{\hat \Omega }}'_\pi} \right.} \right) = 0, \text{ if } i \ne j.
\end{equation}
The quantities in (\ref{partial_Q_partial_theta})--(\ref{partial2_Q_partial_tau_i_partial_tau_j}) are all evaluated at ${\bf{\Omega }}_\pi ^{(t)}$. Repeating the calculation of (\ref{Newton_update}) until $\{{\bf{\hat \Xi}}^{(t)}\}$ converges, the limit point ${\bf{\hat \Xi}}$ of $\{{\bf{\hat \Xi}}^{(t)}\}$ is the solution for (\ref{Update_Xi_condition}), and also the updated estimate of ${\bf{\Xi}}$.

The convergence of the EM algorithm is guaranteed and the detailed analysis can be found in \cite{dempster1977maximum, wu1983convergence}, that is to say, by iteratively alternating between \emph{E-step} and \emph{M-step}, a locally optimal solution for (\ref{Optimization_Identification_Estimation_1}) can be obtained. It is worth mentioning that since we do not require a very accurate solution for the relaxed optimization problem in (\ref{Optimization_Identification_Estimation_1}), once the difference between the updated and current estimates is sufficiently small, we can terminate the iterations in the EM algorithm and utilize the current estimate of ${{{\bf{\Omega }}_\pi }}$ in the following rounding step.    

\subsection{Constrained Variable Threshold Rounding and Barrier Method}
\label{Section_Rounding_Barrier}

By utilizing the EM algorithm as illustrated in \emph{Section \ref{Sec_Random_Relaxation}}, we can obtain the solution $\bf{\hat \Omega}_\pi$ for the relaxed optimization problem in (\ref{Optimization_Identification_Estimation_1}). The element ${\hat \pi}_j$ of $\bf{\hat \Omega}_\pi$ can be interpreted as the probability of the $j$-th sensor being attacked over time. However, according to (\ref{Optimization_Constraint_Sum_Ij}) and (\ref{Definition_pi_j}), we know that before relaxation, ${\hat \pi}_j \in \{0,1\}$ and ${\bf{1}}^T{\boldsymbol{\hat \pi}} < N/2$. To this end, we consider the task of rounding ${\boldsymbol{\hat \pi}}$ to a valid binary vector. To accomplish this task, we propose a constrained variable threshold rounding (CVTR) approach which is based on the heuristic developed by Zymnis \emph{et al.} \cite{zymnis2009relaxed}. The basic idea of the CVTR is that we first round ${\boldsymbol{\hat \pi}}$ to generate a set of most likely probability vectors $\{ {{{{\boldsymbol{\tilde \pi }}}^{(l)}}} \}$ with binary elements which satisfy the constraints in (\ref{Optimization_Constraint_Sum_Ij}). Then, under constraint (\ref{Optimization_Constraint_tau_j}), the joint maximum likelihood estimate of the desired vector parameter and attack vector parameters are pursued over the generated set of valid probability binary vectors $\{ {{{{\boldsymbol{\tilde \pi }}}^{(l)}}} \}$.

We first generate the set of the most likely valid binary probability vectors $\{ {{{{\boldsymbol{\tilde \pi }}}^{(l)}}} \}$ by employing the CVTR which can be described as
\begin{align} \notag
\left\{ {{{{\boldsymbol{\tilde \pi }}}^{(l)}}} \right\} \buildrel \Delta \over = & \bigg\{ {{\mathop{\rm sgn}} \left( {{\boldsymbol{\hat \pi }} - \lambda {\bf{1}}} \right):  0 \le \lambda  \le 1,} \bigg. \\ \label{Definition_rounding_pi_j}
& \qquad  \qquad  \qquad \quad \left. {{{\left\| {{\mathop{\rm sgn}} \left( {{\boldsymbol{\hat \pi }} - \lambda {\bf{1}}} \right)} \right\|}_1} < \frac{N}{2}} \right\}.
\end{align}
Since the $j$-th element ${\tilde \pi}_j^{(l)}$  of ${{{{\boldsymbol{\tilde \pi }}}^{(l)}}}$ denotes the probability the $j$-th sensor is attacked, each probability vector ${{{{\boldsymbol{\tilde \pi }}}^{(l)}}}$ with binary values corresponds to a deterministic state variable vector ${\boldsymbol{\tilde \eta}}^{(l)}$ as following
\begin{equation}
{\boldsymbol{\tilde \eta}}^{(l)} = {{{{\boldsymbol{\tilde \pi }}}^{(l)}}}, \; \forall l.
\end{equation}
We refer to $\{{\boldsymbol{\tilde \eta}}^{(l)}\}$ as the set of the most likely state variable vectors, and we 
only consider the combinations in this set. Further, it is seen from (\ref{Definition_rounding_pi_j}) that as $\lambda$ increases from $0$ to $1$, this approach only generates up to $\left\lfloor {N/2} \right\rfloor$ distinct valid binary probability vectors.
Thus, it is feasible to exhaustively evaluate the maximum likelihood function, which is maximized with respect to $\bf{\Xi}$, for each given ${\boldsymbol{\tilde \eta}}^{(l)}$. As a result, the optimization problem in (\ref{Optimization_Identification_Estimation}) can be reduced to 
\begin{subequations} \label{Optimization_Identification_Estimation_Rounding}
\begin{align} \label{Optimization_Identification_Estimation_Rounding_Object}
{{{\bf{\hat \Omega }}}_R}  = &  \left[ {{{{\bf{\hat \Xi }}}_R^T},{{{\boldsymbol{\hat \eta }}}_R^T}} \right]^T \! \!=  \arg \mathop {\max }\limits_{{\boldsymbol{\eta }} \in \left\{ {{{{\boldsymbol{\tilde \eta }}}^{(l)}}} \right\}} \mathop {\max }\limits_{\bf{\Xi }} L\left( {\bf{\Omega}} \right)  \\ \label{Optimization_Rounding_Constraint_tau_j}
&  \quad  \text{s. t. } \frac{1}{{{K_j}}}\sum\limits_{k = 1}^{{K_j}} {{{\left\| {{\bf{\tilde q}}_j^{(k)} - {\bf{q}}_j^{(k)}} \right\|}_2}}  \ge {d_q}, \forall {{\eta}_j} = 1.
\end{align}
\end{subequations}


As (\ref{Optimization_Identification_Estimation_Rounding}) demonstrates, we need to solve the inner maximization for each candidate state variable vector ${\boldsymbol{\tilde \eta}}^{(l)}$, and then keep the solution which gives rise to the maximal objective function in (\ref{Optimization_Identification_Estimation_Rounding}). Noticing that the constraint in (\ref{Optimization_Rounding_Constraint_tau_j}) only has effects on the inner maximization, the inner constrained maximization for each ${\boldsymbol{\tilde \eta}}^{(l)}$ in (\ref{Optimization_Identification_Estimation_Rounding}) can be converted to an unconstrained problem by employing a logarithmic barrier function as
\begin{align} \notag
\mathop {\max }\limits_{\bf{\Xi }} &  \left\{ {\sum\limits_{j = 1}^N {\sum\limits_{k = 1}^{{K_j}} {\left[ {{\tilde \eta }_j^{(l)}\ln {{\tilde q}_{j{r_{jk}}}^{(k)}} + \left( {1 - {\tilde \eta }_j^{(l)}} \right)\ln {q_{j{r_{jk}}}^{(k)}}} \right]} } } \right. \\ \label{Optimization_barrier_inner_problem}
&   \!+\! \left. {\mu \sum\limits_{j = 1}^N {{\tilde \eta }_j^{(l)}\ln \left( \frac{1}{{{K_j}}}\sum\limits_{k = 1}^{{K_j}} {{{\left\| {{\bf{\tilde q}}_j^{(k)} - {\bf{q}}_j^{(k)}} \right\|}_2}} \!\! -  {d_q} \right)} } \right\},
\end{align}
where the positive barrier parameter $\mu$ determines the accuracy with which (\ref{Optimization_barrier_inner_problem}) approximates the inner constrained maximization in (\ref{Optimization_Identification_Estimation_Rounding}). Since the objective function in (\ref{Optimization_barrier_inner_problem}) is differentiable, the unconstrained problem in (\ref{Optimization_barrier_inner_problem}) can be similarly solved by Newton's Method as in \emph{Section \ref{Section_Updated_Xi}} for any given $\mu$. 

Let ${{{\bf{\hat \Xi }}}_\mu ^{(l)}}$ denote the solution of (\ref{Optimization_barrier_inner_problem}) for any given ${\boldsymbol{\tilde \eta}}^{(l)}$ and $\mu$, and let $L_*^{(l)}$ represent the optimal objective value of the inner constrained maximization in (\ref{Optimization_Identification_Estimation_Rounding_Object}) for any given ${\boldsymbol{\tilde \eta}}^{(l)}$. It can be shown that as $\mu  \to 0$, any limit point ${{{\bf{\hat \Xi }}}_* ^{(l)}}$ of the sequence ${\{ {{{\bf{\hat \Xi }}}_\mu ^{(l)}}\}_\mu }$ is a solution of the inner constrained maximization in (\ref{Optimization_Identification_Estimation_Rounding}) \cite{nash1994numerical}. 
Thus, we can obtain an accurate solution of the inner constrained maximization in (\ref{Optimization_Identification_Estimation_Rounding}) by iteratively solving (\ref{Optimization_barrier_inner_problem}) for a sequence $\{{\mu}_m\}$ of positive barrier parameters, which decrease monotonically to zero, such that the solution ${{{\bf{\hat \Xi }}}_{\mu_m} ^{(l)}}$ for ${\mu}_m$ is chosen as the starting point for the next iteration with barrier parameter ${\mu}_{m+1}$. By defining ${l^ * } \buildrel \Delta \over = {\max _l} \; L_ * ^{(l)}$, the solution of the constrained optimization problem in (\ref{Optimization_Identification_Estimation_Rounding}) can be obtained as
\begin{equation} \label{final_estimate}
{{{\bf{\hat \Omega }}}_R} = {\left[ {{{{\bf{\hat \Xi }}}_R},{{{\boldsymbol{\hat \eta }}}_R}} \right]^T} = {\left[ {{{\left( {{\bf{\hat \Xi }}_*^{(l^*)} } \right)}^T},{{\left( {{{\boldsymbol{\tilde{\eta} }}^{(l^*)}}} \right)}^T}} \right]^T}.
\end{equation}

\subsection{Discussion}


It is well known that the condition number of the Hessian matrix of the logarithmic barrier function in (\ref{Optimization_barrier_inner_problem}) might become increasingly larger as the barrier parameter decreases to $0$. In order to overcome the ill-conditioning issue in practical computation, the numerically stable approximation of the Newton direction can be utilized in Newton's method for solving (\ref{Optimization_barrier_inner_problem}) with small barrier parameter, see \cite{nash1994numerical} and the references therein. It is worth mentioning that to preserve the generality, we don't make additional assumptions to ensure the convexity of the objective functions in the section. Hence, the EM algorithm and Newton's method involved in our approach might converge to a locally optimal point if the starting point is not close to the globally optimal point. To avoid this possibility, multiple starting points can be employed and we choose the one that yields the maximal objective function at convergence \cite{kay1993fundamentals}.


The proposed approach in (\ref{Optimization_Identification_Estimation_Rounding})--(\ref{final_estimate}) attempts to use all sensor
data, whether attacked or not, in an attempt to optimize the described
objective function.  Thus, for some scenarios where the spoofing attacks
are not OGDSAs, similar to \cite{zhang2015Asymptotically}, one expects the proposed approach will
outperform
the estimation approach which only utilizes the unattacked data to estimate the
desired vector parameter. For example, please refer to the numerical results in \emph{Section \ref{Section_Data_injection_attacks_in_MIMO_radars}}.


\section{Numerical Results} 
\label{Section_Numerical_Results}

In this section, we investigate the performance of the approaches proposed in \emph{Section \ref{Sec_JointIdentification}} for some practical cases. The numerical results show that under OGDSAs, the approaches proposed in \emph{Section \ref{Sec_JointIdentification}} are not able to obtain performance that is better than the optimal performance which ignores the attacked sensors, which is consistent with the theory described in \emph{Section \ref{Section_Optimality_Spoofing_Attacks}}.

\subsection{DRFM Attacks in MIMO Radars} 
\label{Section_DRFM_Attacks}

First, we consider MIMO radar with $1$ transmit station and $N=10$ moderately spaced receive stations under the spoofing attack using a DRFM in a generalization of (\ref{delay_receive_model})--(\ref{DRFM_pdf}). 
The first $3$ receive stations are under attack. Each station makes $M$ measurements of each pulse in the pulse train, and employs an identical $4$-bit quantizer with a set of thresholds $\{-\infty,-5,-4,...,9,\infty\}$ to convert analog measurements to quantized data before transmitting them to the FC. Without any attack, the $m$-th measurement of the $k$-th pulse in the pulse train at the $j$-th station can be expressed as
\begin{equation}
x_{jm}^{(k)} = \sqrt {{E_j}} {a_j}s\left( {t_{jm}^{(k)} - \theta_j } \right) + n_{jm}^{(k)},
\end{equation}
where $\theta_j$ is the desired parameter, $m=1,2,...,M$, $k=1,2,...,K$, and $K$ is the total number of pulses in the pulse train. Similar to (\ref{DRFM_after_attack_measurement}), if the $j$-th station is under attack, the $m$-th measurement of the $k$-th pulse in the pulse train is
\begin{equation} \label{after_attack_DRFM}
\tilde x_{jm}^{(k)} = \sqrt {{E_j}} {a_j}s\left( {t_{jm}^{(k)} - {\theta _j} - {\xi _j}} \right) + n_{jm}^{(k)},
\end{equation}
where ${\xi _j}$ is the delay introduced by the DRFM.

\begin{figure}[htb]
	\vspace{0.04in}
	\centering
	\centerline{
		\includegraphics[width=0.40\textwidth]{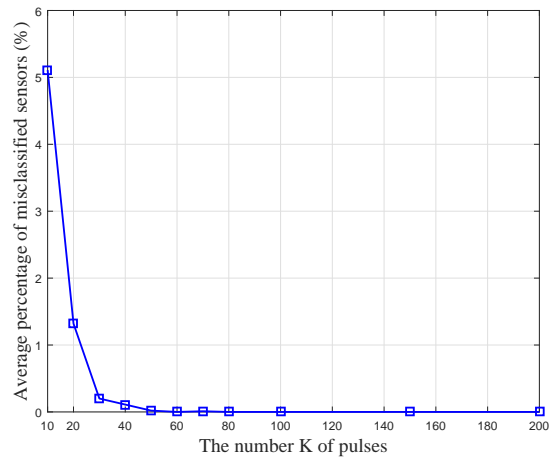}
	}
	\caption{Performance of identifying the DRFM attacks.}
	\label{Identification_Delay_fig}
\end{figure}

Assume $\{n_{jm}^{(k)}\}$ is an independent and identically distributed zero-mean Gaussian noise sequence with variance $\sigma ^2$.
The signal $s\left( t \right)$ is a Gaussian pulse signal \cite{he2010noncoherent}, that is,
\begin{equation} \label{Gaussian_pulse_signal}
s\left( t \right) = {\left( {\frac{2}{{{T^2}}}} \right)^{\frac{1}{4}}}\exp \left( { - \frac{{\pi {t^2}}}{{{T^2}}}} \right),
\end{equation}
and the sampling times are $t_{jm}^{(k)} = (m - 1)\Delta t$, $\forall m=1,2,...,M$. Moreover, we assume the distance between the target and any receiving station is much larger than the distance between every pair of stations, and hence, we can assume $\theta_j = \theta$ for all $j$. In the simulations, let $T=0.1$, $\Delta t = 0.001$, $\theta=0.02$, $M=3$, $\sigma^2 =5$, and $E_j = 1$, $a_j = 1$ for all $j$. In addition, the values of the attack parameters are $\xi_1 = 0.04$, $\xi_2 = 0.05$, $\xi_3 = 0.06$, and the threshold defined in \emph{Assumption \ref{Assumption_Significant_Attack}} in (\ref{impact_assumption_pmf}) is $d_q = 0.15$. We first test the performance of the approach which employs the random relaxation (RR) with the EM and CVTR in identifying the attacked and unattacked sensors. Fig. \ref{Identification_Delay_fig} illustrates the Monte Carlo approximation ($1000$ times) of the ensemble average of the percentage of all mis-classified sensors as a function of the number $K$ of pulses in the pulse train. As Fig. \ref{Identification_Delay_fig} shows, the average percentage of mis-classified sensors decreases towards $0$ as $K$ increases. 

\begin{figure}[htb]
	\centering
	\centerline{
		\includegraphics[width=0.40\textwidth]{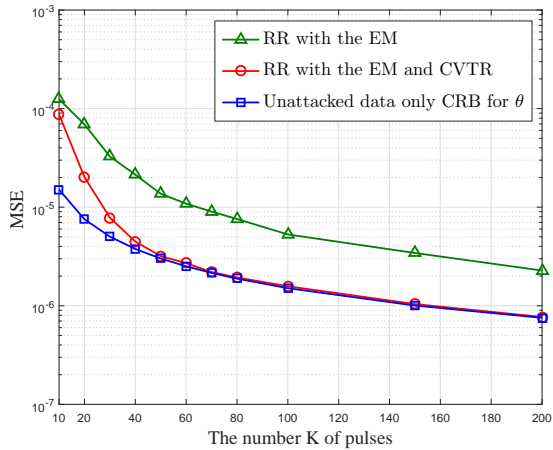}
	}
	\caption{Estimation performance of the proposed approaches under DRFM attacks.}
	\label{EstimationPeformance_Delay_fig}
\end{figure}

Next, we examine the estimation performance of the proposed approaches in \emph{Section \ref{Sec_JointIdentification}}, that is, the approach which employs the RR with the EM, and the approach that employs the RR with the EM and CVTR. Fig. \ref{EstimationPeformance_Delay_fig} depicts the mean squared error (MSE) performance of the two approaches for estimating $\theta$ on a log scale. For comparison, in Fig. \ref{EstimationPeformance_Delay_fig}, the CRB for $\theta$ which knows which sensors are attacked and only uses data from unattacked sensors is also provided\footnote{It is worth mentioning that the CRB for $\theta$ which knows which sensors are attacked and only uses data from unattacked sensors is equal to the CRB for estimating $\theta$ under ISA. Hence, the blue curve marked with squares in Fig.3 also indicates the CRB performance under ISA.}.
It is seen that as $K$ increases, the MSE performance of the approach with CVTR for estimating $\theta$ converges to the CRB for $\theta$ which knows which sensors are attacked and only uses data from unattacked sensors.
The large $K$ results in Fig. \ref{EstimationPeformance_Delay_fig} also corroborates the previous theoretical results in \emph{Section \ref{Section_Optimality_Spoofing_Attacks}} that under OESAs, 
the attacked data are not useful to reduce the CRB.
In addition, the MSE performance of the approach with CVTR is shown to be better than the approach which only employs the RR with the EM algorithm, which implies that the proposed constrained variable threshold rounding can further improve the estimation performance for estimating the deisred parameter. 

\subsection{Data-injection Attacks in Sensor Networks}
\label{Section_Data_injection_attacks_in_sensor_networks}



Next consider the specific attacks on the range-based localization system described in \cite{Lee2012Characterization}. The attackers modify the receivers to alter the received signal strength to confuse the localization. Consider a case with $N=10$ closely spaced sensors. Each sensor makes $K$ measurements of the physical phenomenon, and employs an identical $4$-bit quantizer with a set of thresholds $\{ {0, \pm 1, \pm 2,..., \pm 7}, \pm\infty \}$ to convert analog measurements to quantized data before transmitting them to the FC. 
From \cite{Lee2012Characterization}, the received signal strength before attack is
\begin{equation} \label{numerical_sensor_network}
{x_{jk}} = \theta  + {n_{jk}}, \; \forall k \text{ and } \forall j,
\end{equation} 
where $\theta$ is a deterministic unknown parameter, and $\{n_{jk}\}$ is an i.i.d. zero-mean Gaussian noise sequence with distribution ${\cal N}(0, \sigma ^2)$. 
 Here, we estimate $\theta$ which allows us to directly calculate the common distance to the emitter, due to a one-to-one relationship.  Further, we assume that the first $3$ sensors in the sensor network are under data-injection spoofing attacks. The after-attack measurements are described as 
\begin{equation} \label{numerical_data_injection}
{{\tilde{x}}_{jk}} = \theta + a_{jk}  + {n_{jk}}, \; \forall k \text{ and } \forall j=1,2,3,
\end{equation}
where $a_{jk}$ is the unknown attack injected at the $j$-th sensor at time $k$. 

\begin{figure}[htb]
	\vspace{0.04in}
	\centerline{
		\includegraphics[width=0.40\textwidth]{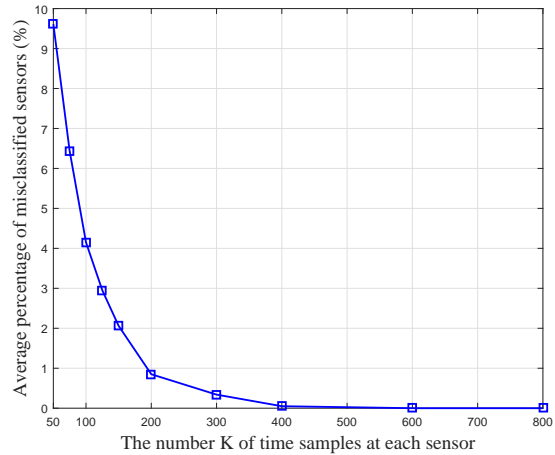}
	}
	\caption{Performance of identifying the data-injection attacks.}
	\label{Identification_Vector_fig}
\end{figure}


\begin{figure}[htb]
	\centerline{
		\includegraphics[width=0.40\textwidth]{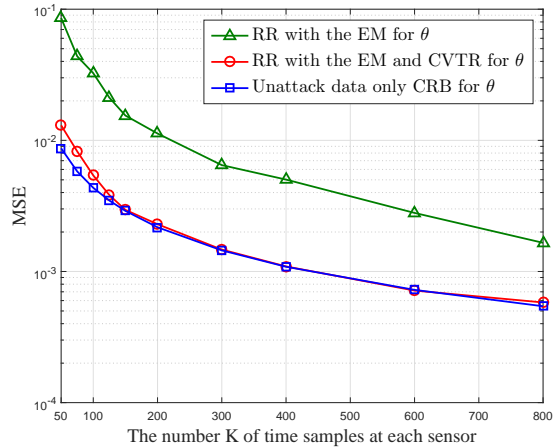}
	}
	\caption{Estimation performance of the proposed approaches for estimating $\theta$.}
	\label{EstimationPeformance_vector_theta_fig}
\end{figure}
\begin{figure}[htb]
	\centerline{
		\includegraphics[width=0.40\textwidth]{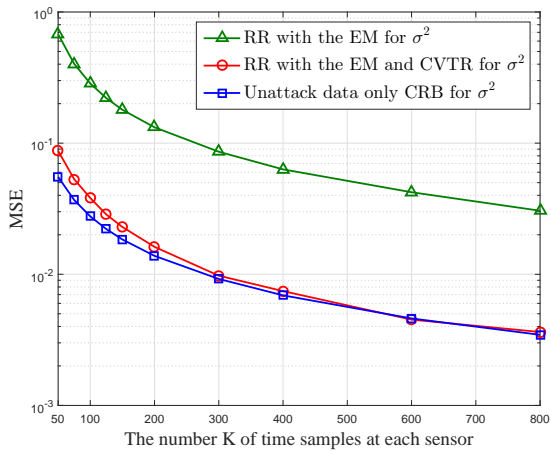}
	}
	\caption{Estimation performance of the proposed approaches for estimating $\sigma^2$.}
	\label{EstimationPeformance_vector_sigma2_fig}
\end{figure}

We consider the scenario that $\theta$ and $\sigma ^2$ are both the parameters of interest. Moreover, the unknown injected attacks $\{{a_{jk}}\}$ are independent random variables, where $a_{jk}$ obeys the Gaussian distribution ${\cal N}(\alpha_j, \beta_j)$ for all $k$. In the simulations, the desired vector parameter ${\boldsymbol{\theta}}  \buildrel \Delta \over =  [\theta, \sigma^2]^T$ and the attack vector parameters ${\{ {{{\boldsymbol{\xi }}^{(j)}} \buildrel \Delta \over = {{[{\alpha _j},{\beta _j}]}^T}} \}_{j = 1,2,3}}$ are ${\boldsymbol{\theta }} = {[1,3]^T}$, ${{\boldsymbol{\xi }}^{(1)}}  = {[ - 1.5, 1]^T}$, ${{\boldsymbol{\xi }}^{(2)}} = {[ -2, 2]^T}$, and ${{\boldsymbol{\xi }}^{(3)}}  = {[1.5, 1]^T}$. The threshold defined in \emph{Assumption \ref{Assumption_Significant_Attack}} in (\ref{impact_assumption_pmf}) is $d_{q} = 0.04$. The performance of the approach which employs the RR with the EM and CVTR in identifying the attacked and unattacked sensors is illustrated in Fig. \ref{Identification_Vector_fig}. Fig. \ref{Identification_Vector_fig} depicts the Monte Carlo approximation ($1000$ times) of the ensemble average of the percentage of all mis-classified sensors versus the number $K$ of time samples at each sensor. It is seen from Fig. \ref{Identification_Vector_fig} that the average percentage of mis-classified sensors reduces towards $0$ as $K$ increases. In Fig. \ref{EstimationPeformance_vector_theta_fig} and Fig. \ref{EstimationPeformance_vector_sigma2_fig}, we plot the MSE performance of our proposed approaches for jointly estimating $\theta$ and $\sigma^2$ on a log scale. The CRBs for estimating $\theta$ and $\sigma^2$ which know which sensors are attacked and only use data from unattacked sensors
are also respectively plotted in Fig. \ref{EstimationPeformance_vector_theta_fig} and Fig. \ref{EstimationPeformance_vector_sigma2_fig} for comparison. As Fig. \ref{EstimationPeformance_vector_theta_fig} and Fig. \ref{EstimationPeformance_vector_sigma2_fig} show, the MSEs of the approach which employs the RR with the EM and CVTR for jointly estimating $\theta$ and $\sigma^2$ respectively converge to the corresponding CRBs which know which sensors are attacked and only uses data from unattacked sensors, and outperform the approach which only employs the RR with the EM algorithm. The large $K$ results in Fig. \ref{EstimationPeformance_vector_theta_fig} and Fig. \ref{EstimationPeformance_vector_sigma2_fig} again justify the previous theoretical results in \emph{Section \ref{Section_Optimality_Spoofing_Attacks}} that under OESAs, 
attacked data is not useful to reduce the CRB.

Note that the model in (\ref{numerical_sensor_network}) is the most studied model in the sensor network estimation literature, typically employed for nonlocalization applications. Thus, the analysis is useful in these other applications also.

\subsection{Data-injection Attacks in MIMO Radars}
\label{Section_Data_injection_attacks_in_MIMO_radars}

It is worth mentioning that the data-injection attacks are OGDSAs for the specific estimation problem described in \emph{Section \ref{Section_Data_injection_attacks_in_sensor_networks}}, but they may not satisfy the necessary and sufficient conditions for an OGDSA for other estimation problems. To demonstrate this, we consider the data-injection attacks in the time delay estimation problem described in \emph{Section \ref{Section_DRFM_Attacks}}. To be specific, if the $j$-th station is under a data-injection attack, rather than (\ref{after_attack_DRFM}), the $m$-th after-attack measurement of the $k$-th pulse in the pulse train is
\begin{equation} \label{Data_injection_attack_time_delay}
\tilde x_{jm}^{(k)} = \sqrt {{E_j}} {a_j}s\left( {t_{jm}^{(k)} - {\theta _j}} \right) + {\xi _j} + n_{jm}^{(k)},
\end{equation}
where the signal $s(t)$ is defined in (\ref{Gaussian_pulse_signal}), $\{n_{jm}^{(k)}\}$ is an independent and identically distributed zero-mean Gaussian noise sequence with variance $\sigma ^2$, and the sampling times are $t_{jm}^{(k)} = (m - 1)\Delta t$, $\forall m=1,2,...,M$. In the following simulations, the values of most of system parameters are set to be the same as those in \emph{Section \ref{Section_DRFM_Attacks}}, except that $M=40$ and the values of the attack parameters are $\xi_1 = 1$, $\xi_2 = -2$ and $\xi_3 = -1$. In addition, the threshold defined in \emph{Assumption \ref{Assumption_Significant_Attack}} in (\ref{impact_assumption_pmf}) is chosen to be $d_q = 0.075$. 

\begin{figure}[htb]
	\vspace{0.04in}
	\centerline{
		\includegraphics[width=0.40\textwidth]{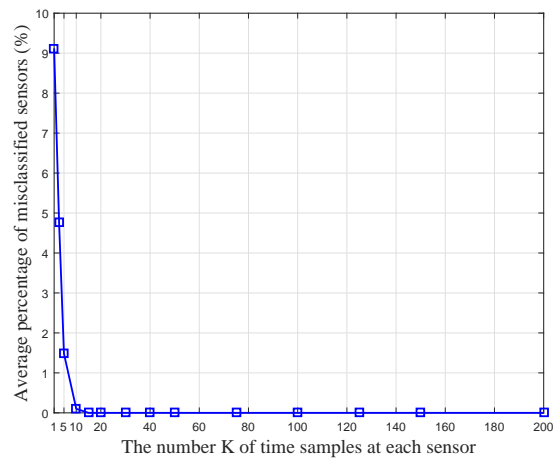}
	}
	\caption{Performance of identifying the data-injection attacks in the time delay estimation problem.}
	\label{Identification_DI_MIMO_fig}
\end{figure}

\begin{figure}[htb]
	\centerline{
		\includegraphics[width=0.40\textwidth]{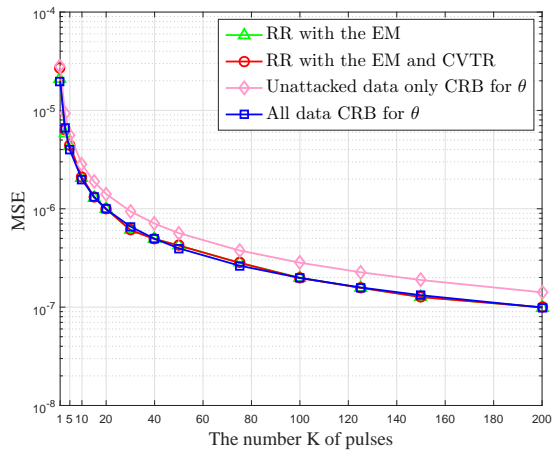}
	}
	\caption{Estimation performance of the proposed approaches for estimating $\theta$ in the presence of the data-injection attacks in the time delay estimation problem.}
	\label{EstimationPeformance_DI_MIMO_fig}
\end{figure}

We first examine the performance of the proposed approach in identifying the attacked and unattacked sensors in the presence of data-injection attacks in the time delay estimation problem. Fig. \ref{Identification_DI_MIMO_fig} depicts the Monte Carlo approximation ($500$ times) of the ensemble average of the percentage of all mis-classified sensors versus the number $K$ of pulses in the pulse train. It is seen that the average percentage of mis-classified sensors reduces to $0$ very rapidly as $K$ increases. Fig. \ref{EstimationPeformance_DI_MIMO_fig} presents the MSE performance of the proposed estimation approaches for estimating $\theta$ plotted on a log scale. The CRB for estimating $\theta$ which knows which sensors are attacked and uses data from all sensors is also plotted in Fig. \ref{EstimationPeformance_DI_MIMO_fig} along with the CRB for estimating $\theta$ which knows which sensors are attacked but only uses data from unattacked sensors.  
Fig. \ref{EstimationPeformance_DI_MIMO_fig} shows that the CRB which uses only the unattacked sensor data is strictly larger than the CRB which uses all the data. In an OGDSA, the data at the attacked sensor cannot be used to improve performance as stated in \emph{Section \ref{Section_Optimality_Spoofing_Attacks}}. Thus the results in Fig. \ref{EstimationPeformance_DI_MIMO_fig} illustrate that the data-injection attack is not OGDSA for the time delay estimation problem in (\ref{Data_injection_attack_time_delay}). This can also be verified using our theory. Interestingly, each of the EM-based approaches we described provides an MSE very close to the CRB from using all data, as shown in Fig. \ref{EstimationPeformance_DI_MIMO_fig}.


\section{Conclusion}
\label{SEC:Conclusion}

In this paper, we study the distributed estimation of a deterministic vector parameter by using quantized data in the presence of spoofing attacks. A generalized attack model is employed which manipulates the data using transformations with arbitrary functional forms determined by some attack parameters whose values are unknown to the attacked system.
Novel necessary and sufficient conditions are provided under which these transformations provide an OGDSA. It is shown that an OGDSA implies that either the FIM under the conditions of \emph{Definition \ref{Definition_Opt_Measurement_Attacks}} for jointly estimating the desired and attack parameters is singular or that the attacked system is unable to improve the CRB under the conditions of \emph{Definition \ref{Definition_Opt_Measurement_Attacks}} for the desired vector parameter 
even though the joint FIM is nonsingular. 
It is demonstrated that it is always possible to construct an OGDSA by properly employing a sufficiently large dimension  attack vector parameter relative to the number of quantization levels employed, which was not observed previously.
In addition, we demonstrate that under the conditions of \emph{Definition \ref{Definition_Opt_Measurement_Attacks}}, a spoofing attack can corrupt the original measurements to make them useless in terms of reducing the CRB for estimating the desired vector parameter if and only if it is an OGDSA. 
In order to illustrate the theory in a concrete way, we also provide some numerical results considering some OGDSAs.  For a specific class of OGDSAs, an enhanced  EM-based algorithm that attempts to use all the attacked and unattacked data to jointly estimate the desired and attack parameters is shown, for a sufficient number of observations, to essentially achieve the CRB which knows which sensors are attacked and only uses data from unattacked sensors. This tallies with the theoretical results that the attacked data is not useful under an OGDSA.  For completeness, we specify the EM-based algorithm for general attacks and enhance it with a heuristic rounding approach previously suggested by others in a different application which seems to significantly improve the EM-based algorithm.

\bibliographystyle{IEEEtran}
\bibliography{Attack}

\end{document}